\documentclass[11pt]{article}
\pdfoutput=1
\usepackage{soul}
\usepackage{jheppub}
\usepackage{amsfonts}
\usepackage{amsmath}
\allowdisplaybreaks[4]         
\usepackage{amssymb}   
\usepackage{euscript}       
\usepackage{color}          
\usepackage{tensor}        
\usepackage{comment}

\newcommand{\req}[1]{(\ref{#1})} 

\newcommand{\bea}{\begin{eqnarray}}
\newcommand{\eea}{\end{eqnarray}}
\newcommand{\ba}{\begin{eqnarray}}
\usepackage{braket}
\newcommand{\ea}{\end{eqnarray}}

\newcommand{\beq}{\begin{equation}}
\newcommand{\eeq}{\end{equation} }
\newcommand{\beqa}{\begin{eqnarray}}

\newcommand{\eeqa}{\end{eqnarray}}
\newcommand{\beqar}{\begin{eqnarray*}}
\newcommand{\eeqar}{\end{eqnarray*}}

\newcommand{\be}{\begin{equation}}
\newcommand{\ee}{\end{equation}}

\newtheorem{conjecture}{Conjecture}

\newcommand{\ele}{\mathcal{L}}

\title{ \boldmath Duality-invariant extensions of Einstein-Maxwell theory}

\author[a]{Pablo A. Cano}
\author[b,c]{\& \'Angel Murcia}

\affiliation[a]{Instituut voor Theoretische Fysica, KU Leuven.\\
Celestijnenlaan 200D, B-3001 Leuven, Belgium \vspace{0.1cm}}
\affiliation[b]{Instituto de F\'isica Te\'orica UAM/CSIC. \\
C/ Nicol\'as Cabrera, 13-15, C.U. Cantoblanco, 28049 Madrid, Spain\vspace{0.1cm}}
\affiliation[c]{Department of Mathematics, University of Hamburg. \\ Bundesstra$\beta$e 55, D-20146 Hamburg, Germany}

\emailAdd{pabloantonio.cano@kuleuven.be}
\emailAdd{angel.murcia@csic.es}

\date{\today}
\abstract{
We investigate higher-derivative extensions of Einstein-Maxwell theory that are invariant under electromagnetic duality rotations, allowing for non-minimal couplings between gravity and the gauge field. 
Working in a derivative expansion of the action, we characterize the Lagrangians giving rise to duality-invariant theories up to the eight-derivative level, providing the complete list of operators that one needs to include in the action.  We also characterize the set of duality-invariant theories whose action is quadratic in the Maxwell field strength but which are non-minimally coupled to the curvature. Then we explore the effect of field redefinitions and we show that, to six derivatives, the most general duality-preserving theory can be mapped to Maxwell theory minimally coupled to a higher-derivative gravity containing only  four non-topological higher-order operators.  We conjecture that this is a general phenomenon at all orders, \textit{i.e}, that any duality-invariant extension of Einstein-Maxwell theory is perturbatively equivalent to a higher-derivative gravity minimally coupled to Maxwell theory. Finally, we study charged black hole solutions in the six-derivative theory and we investigate additional constraints on the couplings motivated by the weak gravity conjecture. 

 }

\begin{document} 
\maketitle
\flushbottom

\section{Introduction}
\label{sec:Introduction}

The idea that the laws of nature must be invariant under certain transformations is one of the most fundamental principles of modern physics. In this way, the standard model of particle physics is based on Lorentz symmetry and gauge invariance, while the foundations of General Relativity lie on diffeomorphism invariance. 

The existence of symmetries that persist at the quantum level is a especially powerful tool in the context of Effective Field Theory, since they allow one to reduce the number of operators that can appear in the effective action. For instance, string theory compactified on a torus is invariant under T-duality, which implies the invariance of the lower-dimensional stringy effective actions under the group $\mathrm{O}(d,d)$. This symmetry is claimed to be preserved at all orders in the higher-derivative $\alpha'$-expansion \cite{Sen:1991zi}, and explicit confirmation of this has been reported for the lowest-order terms \cite{Bergshoeff:1995cg,Meissner:1996sa,Eloy:2020dko,Elgood:2020xwu,Ortin:2020xdm,Codina:2020kvj}. However, the computation of these $\alpha'$ corrections from first principles is a complicated problem, and, instead, it turns out that one can use duality invariance to constrain the higher derivative corrections that appear in the stringy effective actions --- see, \textit{e.g.}, \cite{Marques:2015vua,Baron:2017dvb,Razaghian:2018svg}.
This program is particularly successful in the cosmological context, since in that case it is possible to constrain all-orders $\alpha'$-corrections by imposing duality \cite{Hohm:2019jgu} --- see also, \textit{e.g.}, \cite{Hohm:2019ccp,Krishnan:2019mkv,Bernardo:2019bkz,Nunez:2020hxx}.

In this paper, we intend to study a more basic type of duality which already appears at the level of classical electrodynamics: electromagnetic duality. As is well-known, the vacuum Maxwell equations are invariant under the exchange of electric and magnetic fields, and more generally, under a $\mathrm{SO}(2)$ rotation of them. Such transformation also leaves invariant the Maxwell stress-energy tensor and hence is a symmetry of Einstein-Maxwell theory as well. Duality transformations are also symmetries of $\mathrm{U}(1)^N$ gauge theories coupled to scalar fields \cite{Gaillard:1981rj}. These models are in fact ubiquitous in supergravity and string theory \cite{Henneaux:1988gg,Duff:1989tf,Duff:1990hn,Sen:1992fr,Schwarz:1993vs,deWit:2001pz,Ortin:2015hya}, where electromagnetic duality is a part of the U-duality group \cite{Hull:1994ys}.
Let us also mention that one can define more refined notions of duality invariance by allowing the couplings constants of the theory to transform as well \cite{Hull:1995gk,Lazaroiu:2016iav,Lazaroiu:2020vne}. 
However, here we will restrict to the simple but relevant case of theories with a single vector field coupled to gravity such that the space of solutions is preserved under $\mathrm{SO}(2)$ duality rotations.

When higher-derivative corrections are added to the Einstein-Maxwell action, the invariance under duality rotations is generically broken. Therefore, assuming that electromagnetic duality must be preserved constrains the possible terms that can appear in the effective action. One of the advantages of electromagnetic duality invariance with respect to $\mathrm{O}(d,d)$ symmetry is that we do not need to have additional vector fields to constrain the action.  
On the other hand, electromagnetic duality is only a symmetry of the equations of motion, not of the action, and, in addition, it is non-linearly realized in theories with higher-derivative corrections. Therefore, the study of self-dual theories presents some particular complications. 

In the case of pure electromagnetic theories, it is known that the Maxwell Lagrangian allows for non-linear duality-invariant extensions; these were thoroughly studied in \cite{Gibbons:1995cv}\footnote{See also \cite{Gaillard:1997zr,Gaillard:1997rt} for the case of matter couplings.}. Among these, one finds for instance Born-Infeld theory \cite{Born:1934gh,Schrodinger:1935oqa,BialynickiBirula:1992qj} or the Euler-Heisenberg Lagrangian \cite{Heisenberg:1935qt} describing 1-loop QED corrections to Maxwell theory. Recently, Ref~\cite{Bandos:2020jsw} proposed a new non-linear theory that, besides preserving duality rotations, is also invariant under conformal transformations. See also, \textit{e.g.}, \cite{Gibbons:1995ap,Hatsuda:1999ys,Brace:1999zi,Bunster:2011aw,Chemissany:2011yv,BabaeiVelni:2016qea} for other generalizations.

The study of minimally-coupled self-dual theories is facilitated by the fact that, in that case, the most general gauge-invariant Lagrangian can be formed with only two basic invariants\footnote{As we will argue in section~\ref{sec:dualityrot}, duality invariance prevents the appearance of derivatives of the field strength.}, namely $F_{\mu\nu}F^{\mu\nu}$ and $F_{\mu\nu}\star F^{\mu\nu}$. However, in a general effective extension of Einstein-Maxwell theory one should consider as well the addition of non-minimal couplings between the metric and the field strength, \textit{i.e.}, terms like $F^{\mu\nu}F^{\alpha\beta}R_{\mu\nu\alpha\beta}$. This results in a drastic enlargement of the number of invariants one can include in the action, which in turn increases the difficulty of classifying duality-preserving theories.
Remarkably enough, this problem has not been addressed yet in the literature, so the goal of this paper is to provide a thorough characterization of such theories. 

The presence of non-minimal couplings makes it virtually impossible to obtain exactly invariant Lagrangians, so in this paper we tackle this problem by assuming a derivative expansion of the action. We obtain the conditions on the 4-, 6- and 8-derivative Lagrangians that ensure that the theory is a truncation of a duality-preserving one. 
In addition, we will see that, due to the coupling to gravity, metric field redefinitions acquire a very interesting role in the case of duality-invariant theories. In fact, we show that, to six-derivatives, one can get rid of all the higher-derivative terms involving field strengths in any duality-preserving theory by performing such redefinitions, and we conjecture the same to be true at all orders. 

The paper is organized as follows.
\begin{itemize}
\item  In Section~\ref{sec:dualityrot} we consider a general higher-derivative extension of Einstein-Maxwell theory and we determine the necessary and sufficient conditions on the 4-,6- and 8- derivative Lagrangians in order for the theory to preserve electromagnetic duality.
\item Using the previous result, in Section~\ref{sec:all8d} we construct explicitly the most general duality-invariant theory up to eight derivatives. 
\item In Section~\ref{sec:linear} we focus on the family of theories whose action is quadratic in the Maxwell field strength and which are non-minimally coupled to gravity through a susceptibility tensor $\chi^{\mu\nu\rho\sigma}$. We completely characterize the form of this tensor giving rise to duality-preserving theories and we write the equations of motion in a manifestly $\mathrm{SO}(2)$ invariant form.
\item In Section~\ref{sec:redef} we analyze the effect of metric redefinitions on duality-invariant theories. We show that, to the six-derivative level, all the higher-order terms involving Maxwell field strengths can be removed via field redefinitions. In other words, the most general duality-invariant action can be mapped to a higher-derivative gravity minimally coupled to the Maxwell Lagrangian, and we conjecture the same to hold at higher orders.  We also show that the number of (gravitational) higher-order operators in the six-derivative action can be further reduced to five, of which one is topological. 
\item  In Section~\ref{sec:BHs} we study the charged, static and spherically symmetric black hole solutions of the six-derivative theory. We compute their thermodynamic properties and the corrections to the extremal charge-to-mass ratio. We discuss several additional bounds on the couplings by using the recently proposed mild form of the weak gravity conjecture \cite{Hamada:2018dde,Bellazzini:2019xts}. 
\item  Finally, we conclude in Section~\ref{sec:conclusions}.
\end{itemize}

\section{Duality-invariant actions}\label{sec:dualityrot}
In this section we determine the necessary and sufficient conditions for a higher-derivative theory to be invariant under duality rotations. Thus, let us start by writing a general higher-derivative extension of Einstein-Maxwell theory,
\begin{equation}
S=\frac{1}{16\pi G}\int d^4x\sqrt{|g|}\left[R-F^2+\mathcal{L}\left (g^{\mu\nu},R_{\mu\nu\rho\sigma},\nabla_{\alpha}R_{\mu\nu\rho\sigma},\ldots; F_{\mu\nu},\nabla_{\alpha}F_{\mu\nu},\ldots\right)\right]\, .
\end{equation}
Here 
\begin{equation}
F_{\mu\nu}=2\partial_{[\mu}A_{\nu]}
\end{equation}
is the field strength of the gauge field $A_{\mu}$, and $R_{\mu\nu\rho\sigma}$ is the curvature tensor of the metric $g_{\mu\nu}$. On the other hand, $\mathcal{L}$ represents a general invariant formed out of these quantities and their derivatives, and we will soon assume it allows for a polynomial expansion in terms with an increasing number of derivatives. The equations of motion coming from the variation of this action read
\begin{align}
G_{\mu\nu}=&2T_{\mu\nu}-\frac{1}{\sqrt{|g|}}\frac{\delta (\sqrt{|g|}\mathcal{L})}{\delta g^{\mu\nu}}\, ,\\
0=&\nabla_{\nu}\left(F^{\mu\nu}-\frac{1}{2}\frac{\delta\mathcal{L}}{\delta F_{\mu\nu}}\right)\, ,
\end{align}
where

\begin{equation}\label{eq:MaxTmunu}
T_{\mu\nu}=F_{\mu\alpha}\tensor{F}{_{\nu}^{\alpha}}-\frac{1}{4}g_{\mu\nu}F^2\, ,
\end{equation}
is the Maxwell stress-energy tensor, and 
\begin{equation}
\frac{\delta\mathcal{L}}{\delta F_{\mu\nu}}=\frac{\partial\mathcal{L}}{\partial F_{\mu\nu}}-\nabla_{\alpha}\frac{\partial\mathcal{L}}{\partial\nabla_{\alpha}F_{\mu\nu}}+\nabla_{\beta}\nabla_{\alpha}\frac{\partial\mathcal{L}}{\partial\nabla_{\alpha}\nabla_{\beta}F_{\mu\nu}}-\ldots\, .
\end{equation}
For our purposes, it is convenient to rewrite this system of equations by introducing the dual field strength $H$ as follows

\begin{align}
	\label{RotEE}
	G_{\mu\nu}=&2\hat T_{\mu\nu}+\mathcal{E}_{\mu\nu}\, ,\\
	\label{RotCons}
	\star H_{\mu\nu}=&-F_{\mu\nu}+\frac{1}{2}\frac{\delta\mathcal{L}}{\delta  F^{\mu\nu}}\, ,\\
	\label{RotBi}
	d\begin{pmatrix}
		F\\
		H
	\end{pmatrix}=&0\, ,
\end{align}
where now

\begin{align}\label{eq:hatTmunu}
\hat T_{\mu\nu}&=-F_{\langle \mu|\alpha}\star \tensor{H}{_{|\nu\rangle}^{\alpha}}\, ,\\
\mathcal{E}_{\mu\nu}&=-\frac{1}{\sqrt{|g|}}\frac{\delta (\sqrt{|g|}\mathcal{L})}{\delta g^{\mu\nu}}+F_{\langle \mu|\alpha}\frac{\delta\mathcal{L}}{\delta \tensor{F}{^{|\nu\rangle}_{\alpha}}}\, ,
\label{eq:Emunu1}
\end{align}
and here $\langle\mu\nu\rangle$ represents the symmetric and traceless part of a tensor, this is
\begin{equation}
X_{\langle\mu\nu\rangle}=X_{(\mu\nu)}-\frac{1}{4}g_{\mu\nu}g^{\alpha\beta}X_{\alpha\beta}\, .
\end{equation}
In addition, the Hodge dual is defined by
\begin{equation}
\star H_{\mu\nu}\equiv\frac{1}{2}\epsilon_{\mu\nu\alpha\beta}H^{\alpha\beta}\, ,
\end{equation}
where the Levi-Civita symbol $\epsilon_{\mu\nu\alpha\beta}$ is such that $\epsilon_{0123}=\sqrt{|g|}$.

In this formulation, $F$ and $H$ are taken as independent fundamental variables, and the equations of motion impose that they are closed 2-form and related by the constitutive relation \req{RotCons}.
Let us now analyze if the equations \eqref{RotEE}, \eqref{RotCons} and \eqref{RotBi} have any symmetry. It is obvious that the form of the two Bianchi identities \eqref{RotBi} is preserved if we consider any $\mathrm{GL}(2,\mathbb{R})$ transformation of $F$ and $H$. Now, in the case without corrections, the constraint equation \req{RotCons} is only preserved if such transformation belongs to $\mathbb{R} \times \mathrm{SO}(2)$, while the right-hand-side of Einstein's equations is preserved if the transformation belongs to $\mathrm{SL}(2,\mathbb{R})$. Thus, written in this way, the Einstein-Maxwell system is invariant under the intersection of these two groups, which are the $\mathrm{SO}(2)$ transformations

\begin{equation}\label{FHrotation}
\begin{pmatrix}
F\\
H
\end{pmatrix}=
\begin{pmatrix}
\cos\alpha &\sin\alpha \\
-\sin\alpha&\cos\alpha 
\end{pmatrix}\, 
\begin{pmatrix}
F'\\
H'
\end{pmatrix}\, .
\end{equation}

\noindent
Once the corrections are taken into consideration, however, the equations are generally not invariant under this rotation, so, our goal is to determine which Lagrangians do give rise to duality-invariant equations of motion. 

Let us first start by noticing that, if the Lagrangian $\mathcal{L}$ depends non-degenerately on the derivatives of $F_{\mu\nu}$, then the relation $H(F)$ given by \eqref{RotCons} is differential. This means that the inverse relation $F(H)$ must involve integration. Now imagine, for instance, a rotation with an angle $\alpha=\pi/2$. In that case, the new fields $F'$, $H'$ satisfy the relation

\begin{equation}\label{ConsFprime1}
\star F'_{\mu\nu}=H'_{\mu\nu}-\frac{1}{2}\frac{\delta \mathcal{L}}{\delta  F^{\mu\nu}}\bigg|_{F\rightarrow H'}\, .
\end{equation}
Then, if the equations of motion are invariant under duality rotations, this relation should be equivalent to the one obtained by inverting \eqref{RotCons}, but we see that this is not possible since, as we mentioned, $F(H)$ must involve integration while \req{ConsFprime1} is again differential. Therefore, the conclusion is that the equations of motion cannot be duality invariant if the Lagrangian contains derivatives of the field strength. Thus, duality restricts the set of allowed Lagrangians to be of the form
\begin{equation}
\mathcal{L}\left (g^{\mu\nu},R_{\mu\nu\rho\sigma},\nabla_{\alpha}R_{\mu\nu\rho\sigma},\ldots; F_{\mu\nu}\right)\, .
\end{equation}
It is important to clarify a subtle point, though. If one assumes a perturbative expansion of the Lagrangian, then one may, in fact, invert \eqref{RotCons} perturbatively, yielding a differential relation for $F(H)$. Thus, one can find theories with a differential relation \eqref{RotCons} that are invariant under duality rotations in this perturbative sense  \cite{Chemissany:2011yv}. However, due to the argument above, those theories cannot come from the truncation of a complete theory that is \emph{exactly} invariant. In other words, this means that the summation of the whole perturbative series would not give rise to a sensible theory. Thus, we will restrict ourselves to theories that depend only algebraically on the field strength $F_{\mu\nu}$. 
Let us now study which further constraints duality invariance imposes on the Lagrangian. 

\subsection{Invariance of the constitutive relation}
It is useful to focus first on the constraint equation \req{RotCons}.  After a few algebraic manipulations one can show that the rotated fields $F'$ and $H'$ in \req{FHrotation} satisfy the relation
\begin{equation}\label{ConsFprime}
\star H'_{\mu\nu}=-F'_{\mu\nu}+\frac{1}{2}\hat R\frac{\partial \mathcal{L}}{\partial  F^{\mu\nu}}\bigg|_{F\rightarrow F'\cos\alpha +H' \sin\alpha }\, ,
\end{equation}
where we have introduced the operator

\begin{equation}
\hat R=\cos\alpha+\star\sin\alpha\, .
\end{equation}
Now, duality invariance requires that the transformed fields $F'$ and $H'$ be also a solution of the original equation \req{RotCons}, and this will happen if

\begin{equation}\label{dualcondRot}
\hat R \frac{\partial \mathcal{L}}{\partial F^{\mu\nu}}\bigg|_{F\rightarrow F'\cos\alpha +H' \sin\alpha }=\frac{\partial \mathcal{L}}{\partial F^{\mu\nu}}\bigg|_{F\rightarrow F' }\, .
\end{equation}
Of course, this equality cannot ever hold off-shell since the left-hand-side depends on $H'$, but the right-hand-side does not. However, we only require that both quantities be equal on-shell, which ensures that $F'$ and $H'$ indeed solve the original equation. Thus, we can conveniently write this consistency equation as

\begin{equation}\label{dualcondRot2}
\hat R \frac{\partial \mathcal{L}}{\partial F^{\mu\nu}}\bigg|_{F\rightarrow  F' \cos\alpha+ \star (F'-\frac{1}{2}\frac{\partial \mathcal{L}}{\partial  F})\sin\alpha}=\frac{\partial \mathcal{L}}{\partial F^{\mu\nu}}\bigg|_{F\rightarrow F' }\, .
\end{equation}
This is a highly nonlinear equation that constrains the form of $\mathcal{L}$. In order to make further progress, at this point it is convenient to expand the Lagrangian in a derivative expansion, as 

\begin{equation}
\mathcal{L}=\ell^2\mathcal{L}_{(4)}+\ell^4\mathcal{L}_{(6)}+\ell^6\mathcal{L}_{(8)}+\ldots\, ,
\end{equation}
where $\ell$ is a length scale and each Lagrangian $\mathcal{L}_{(2n)}$ contains $2n$ derivatives of the fields. Then, we will impose duality invariance order by order, but the idea is that the full Lagrangian defines an exactly invariant theory. 
We can solve \req{ConsFprime} perturbatively in $\ell$ and we get 

\begin{align}\notag
\star H'_{\mu\nu}=&-F'_{\mu\nu}+\frac{\ell^2}{2}\hat R\frac{\partial \mathcal{L}_{(4)}}{\partial F^{\mu\nu}}\bigg|_{F\rightarrow \hat R F' }+\ell^4\Bigg [\frac{1}{2}\hat R\frac{\partial \mathcal{L}_{(6)}}{\partial F^{\mu\nu}}  \left. -\frac{1}{4}\hat R  \left(s\star \hat R\frac{\partial \mathcal{L}_{(4)}}{\partial F}\right)^{\alpha\beta} \frac{\partial^2\mathcal{L}_{(4)}}{\partial F^{\alpha \beta} \partial F^{\mu\nu} } \right]\Bigg|_{F\rightarrow \hat R F' }
\\\notag
+&\ell^6\Bigg[\frac{1}{2}\hat R\frac{\partial \mathcal{L}_{(8)}}{\partial F^{\mu\nu}}-\frac{1}{4}\hat R \left(s\star \hat R\frac{\partial \mathcal{L}_{(4)}}{\partial F}\right)^{\alpha\beta} \frac{\partial^2\mathcal{L}_{(6)}}{\partial  F^{\alpha \beta} \partial  F^{\mu\nu} } -\frac{1}{4}\hat R   \left(s\star \hat R\frac{\partial \mathcal{L}_{(6)}}{\delta F}\right)^{\alpha\beta} \frac{\partial^2\mathcal{L}_{(4)}}{\partial  F^{\alpha \beta} \partial  F^{\mu\nu} } \\ 
&+\frac{1}{16}\hat{R}\frac{\partial}{\partial F^{\mu \nu}} \left\lbrace \frac{\partial^2 \mathcal{L}_{(4)}}{\partial F^{\alpha \beta} \partial F^{\rho \sigma}} \left ( s \star \hat{R} \frac{\partial \mathcal{L}_{(4)}}{\partial F} \right)^{\alpha \beta}  \left ( s \star \hat{R} \frac{\partial \mathcal{L}_{(4)}}{\partial F} \right)^{\rho \sigma}  \right\rbrace  \Bigg]\Bigg|_{F\rightarrow \hat R F' }  +\mathcal{O}(\ell^8)\, ,
\label{Hprimeexp}
\end{align}
where  $s\equiv \sin\alpha$. Note that the operator $\hat R$ in the left always acts on the indices $\mu\nu$. 
Now, from \req{dualcondRot} we obtain the following necessary conditions in order for the theory to be invariant under duality rotations (we remove the $F'$ notation for clarity)

 \begin{align}\label{RotalgL4}
\frac{\partial \mathcal{L}_{(4)}}{\partial  F^{\mu\nu}}&=\hat R\frac{\partial \mathcal{L}_{(4)}}{\partial  F^{\mu\nu}}\bigg|_{F\rightarrow \hat R F }\, ,\\
\label{RotalgL6}
\frac{\partial \mathcal{L}_{(6)}}{\partial  F^{\mu\nu}}&=\left[\hat R\frac{\partial \mathcal{L}_{(6)}}{\partial F^{\mu\nu}}  -\frac{1}{2}\hat R  \left(s\star \hat R\frac{\partial \mathcal{L}_{(4)}}{\partial F}\right)^{\alpha\beta} \frac{\partial^2\mathcal{L}_{(4)}}{\partial F^{\alpha \beta} \partial  F^{\mu\nu} } \right]\Bigg|_{F\rightarrow \hat R F }\, ,\\\notag
\frac{\partial \mathcal{L}_{(8)}}{\partial F^{\mu\nu}}&= \Bigg [\hat R\frac{\partial \mathcal{L}_{(8)}}{\partial F^{\mu\nu}}-\frac{1}{2}\hat R \left(s\star \hat R\frac{\partial \mathcal{L}_{(4)}}{\partial F}\right)^{\alpha\beta} \frac{\partial^2\mathcal{L}_{(6)}}{\partial F^{\alpha \beta} \partial  F^{\mu\nu} } -\frac{1}{2}\hat R   \left(s\star \hat R\frac{\partial \mathcal{L}_{(6)}}{\partial F}\right)^{\alpha\beta} \frac{\partial^2 \mathcal{L}_{(4)}}{\partial  F^{\alpha \beta} \partial  F^{\mu\nu} } \\
&+\frac{1}{8} \hat{R}\frac{\partial}{\partial F^{\mu \nu}} \left\lbrace \frac{\partial^2 \mathcal{L}_{(4)}}{\partial F^{\alpha \beta} \partial F^{\rho \sigma}} \left ( s \star \hat{R} \frac{\partial \mathcal{L}_{(4)}}{\partial F} \right)^{\alpha \beta}  \left ( s \star \hat{R} \frac{\partial \mathcal{L}_{(4)}}{\partial F} \right)^{\rho \sigma}  \right\rbrace\Bigg]\Bigg|_{F\rightarrow \hat R F }\,. 
\label{RotalgL8}
\end{align}

\noindent
Let us further investigate the implications of these relations for the Lagrangian. First, notice that each Lagrangian is formed out of monomials that can be schematically denoted by $F^n \nabla^q R^p$. Clearly, duality rotations do not mix terms with different values of $n$, $q$ and $p$, and hence, if the theory preserves duality, the relations above are satisfied by each of these families of monomials independently.  Now, let us note that, for every such monomial we have the identity 

\begin{equation}
F^{\mu \nu} \frac{\partial \mathcal{L}}{\partial F^{\mu \nu}}=n \ele\, ,
\end{equation}
since they are homogeneous functions of degree $n$ in $F$.  Likewise, we have 
\begin{equation}
F^{\mu \nu}\left(\hat R\frac{\partial \mathcal{L}}{\partial  F^{\mu\nu}}\bigg|_{F\rightarrow \hat R F }\right)=n \ele \Big|_{F\rightarrow \hat R F }\, .
\end{equation}
Let us then apply these results to the case of $\ele_{(4)}$.
Since the equation \req{RotalgL4} must hold for every type of monomial, we conclude that the four-derivative Lagrangian must satisfy the condition

\begin{equation}
\label{eq:l4wcd}
\ele_{(4)} ( \hat{R} F)=\ele_{(4)}(F)\, ,
\end{equation}
which can be rephrased as the fact that $\ele_{(4)}$ remains invariant under a rotation of $F$ and $\star F$.  Let us now consider the case of $\mathcal{L}_{(6)}$. First, it is convenient to rewrite Eq.~\req{RotalgL6} as follows

\begin{equation}
\frac{\partial \mathcal{L}_{(6)}}{\partial  F^{\mu\nu}}=\left[\hat R\frac{\partial \mathcal{L}_{(6)}}{\partial F^{\mu\nu}}  -\frac{1}{4}\hat R \frac{\partial}{ \partial  F^{\mu\nu}} \left( \left(s\star \hat R\frac{\partial \mathcal{L}_{(4)}}{\partial F}\right)^{\alpha\beta} \frac{\partial\mathcal{L}_{(4)}}{\partial F^{\alpha \beta} } \right)\right]\Bigg|_{F\rightarrow \hat R F }\, .
\end{equation}

\noindent
Then, proceeding with the same logic as before, we can split this expression into monomials of degree $n$ in $F$, and contracting with $F^{\mu\nu}$ we find that 

\begin{equation}
\label{eq:l6wcd}
\ele_{(6)} ( \hat{R} F)-\ele_{(6)}(F)=\frac{1}{4}  \left(s\star \hat R\frac{\partial \mathcal{L}_{(4)}}{\partial F}\right)^{\alpha\beta} \frac{\partial\mathcal{L}_{(4)}}{\partial F^{\alpha \beta} }\Bigg|_{F\rightarrow \hat R F }\, .
\end{equation}
Now this tells us that $\ele_{(6)}$ does not remains invariant under a rotation of $F$ and $\star F$ since there is an inhomogeneous term associated to $\ele_{(4)}$. Clearly, this can be traced back to the fact that duality is non-linearly realized in the Lagrangian formulation.  Then, the general solution to this equation can be expressed as

\begin{equation}\label{eq:L6dec}
\ele_{(6)}=\mathcal{L}_{(6)}^{\mathrm{H}}+\mathcal{L}_{(6)}^{\mathrm{IH}}\, ,
\end{equation}
where $\mathcal{L}_{(6)}^{\mathrm{H}}$ is the general solution of associated homogeneous equation, and therefore satisfies  
\begin{equation}
\mathcal{L}_{(6)}^{\mathrm{H}} (\hat{R} F)=\mathcal{L}_{(6)}^{\mathrm{H}} (F)\,,
\end{equation}
and $\mathcal{L}_{(6)}^{\mathrm{IH}}$ is a particular solution of the full inhomogeneous equation.  Let us show that a particular solution is given by 

\begin{equation}
\mathcal{L}_{(6)}^{\mathrm{IH}}=-\frac{1}{8} \frac{\partial \mathcal{L}_{(4)}}{\partial F^{\alpha \beta}} \frac{\partial \mathcal{L}_{(4)}}{\partial F_{\alpha \beta}}\, .
\label{eq:l6inh}
\end{equation}
In fact, using the property of $\mathcal{L}_{(4)}$ in \req{RotalgL4}, we have 

\begin{equation}
\mathcal{L}_{(6)}^{\mathrm{IH}}(F)=-\frac{1}{8} \left(\hat R\frac{\partial \mathcal{L}_{(4)}}{\partial F}\right)^{\alpha\beta}\left(\hat R\frac{\partial \mathcal{L}_{(4)}}{\partial F}\right)_{\alpha\beta}\Bigg|_{F\rightarrow \hat R F }=-\frac{1}{8} \left(\hat R^2\frac{\partial \mathcal{L}_{(4)}}{\partial F}\right)^{\alpha\beta}\frac{\partial \mathcal{L}_{(4)}}{\partial F^{\alpha\beta}}\Bigg|_{F\rightarrow \hat R F }\, ,
\end{equation}
and on the other hand, 

\begin{equation}
\mathcal{L}_{(6)}^{\mathrm{IH}}( \hat{R} F)=-\frac{1}{8} \frac{\partial \mathcal{L}_{(4)}}{\partial F^{\alpha \beta}} \frac{\partial \mathcal{L}_{(4)}}{\partial F_{\alpha \beta}}\Bigg|_{F\rightarrow \hat R F }=-\frac{1}{8} \left(\hat R^{-1} \hat R\frac{\partial \mathcal{L}_{(4)}}{\partial F}\right)^{\alpha\beta}\frac{\partial \mathcal{L}_{(4)}}{\partial F^{\alpha\beta}}\Bigg|_{F\rightarrow \hat R F }\, .
\end{equation}
Thus, combining both expressions and using that $\hat R^{-1}=\cos\alpha-\star \sin\alpha$ one easily checks that \req{eq:l6wcd} is satisfied. Let us finally turn to the case of the eight-derivative Lagrangian. After using the decomposition of $\ele_{(6)}$ in \req{eq:L6dec}, we can write the equation \req{RotalgL8} as 

\begin{align}\notag
\frac{\partial \mathcal{L}_{(8)}}{\partial F^{\mu\nu}}&= \Bigg [\hat R\frac{\partial \mathcal{L}_{(8)}}{\partial F^{\mu\nu}}-\frac{1}{2}\hat R\frac{\partial}{\partial F^{\mu\nu}}\left( \left(s\star \hat R\frac{\partial \mathcal{L}_{(4)}}{\partial F}\right)^{\alpha\beta} \frac{\partial\mathcal{L}_{(6)}^{\rm H}}{\partial F^{\alpha \beta} }\right) \\
&+\frac{1}{8} \hat{R}\frac{\partial}{\partial F^{\mu \nu}} \left\lbrace \frac{\partial^2 \mathcal{L}_{(4)}}{\partial F^{\alpha \beta} \partial F^{\rho \sigma}} \left ( s \star \hat{R} \frac{\partial \mathcal{L}_{(4)}}{\partial F} \right)^{\alpha \beta}  \left ( c \hat{R} \frac{\partial \mathcal{L}_{(4)}}{\partial F} \right)^{\rho \sigma}  \right\rbrace\Bigg]\Bigg|_{F\rightarrow \hat R F }\,. 
\end{align}
Then, splitting this expression into monomials and contracting with $F^{\mu\nu}$ we arrive at the equation

\begin{align}\notag
\ele_{(8)} ( \hat{R} F)-\ele_{(8)}(F)&= \Bigg [\frac{1}{2} \left(s\star \hat R\frac{\partial \mathcal{L}_{(4)}}{\partial F}\right)^{\alpha\beta} \frac{\partial\mathcal{L}_{(6)}^{\rm H}}{\partial F^{\alpha \beta} } \\
&-\frac{1}{8} \frac{\partial^2 \mathcal{L}_{(4)}}{\partial F^{\alpha \beta} \partial F^{\rho \sigma}} \left ( s \star \hat{R} \frac{\partial \mathcal{L}_{(4)}}{\partial F} \right)^{\alpha \beta}  \left ( c \hat{R} \frac{\partial \mathcal{L}_{(4)}}{\partial F} \right)^{\rho \sigma} \Bigg]\Bigg|_{F\rightarrow \hat R F }\,. 
\end{align}

\noindent
The general solution can again written as 

\begin{equation}\label{eq:L8dec}
\ele_{(8)}=\mathcal{L}_{(8)}^{\mathrm{H}}+\mathcal{L}_{(8)}^{\mathrm{IH}}\, ,
\end{equation}
where $\mathcal{L}_{(8)}^{\mathrm{H}}$ satisfies

\begin{equation}
\mathcal{L}_{(8)}^{\mathrm{H}} (\hat{R} F)=\mathcal{L}_{(8)}^{\mathrm{H}} (F)
\end{equation}

\noindent
and $\mathcal{L}_{(8)}^{\mathrm{IH}}$ is a particular solution of the full inhomogeneous equation. In this case, one can check that a particular solution is given by 

\begin{equation}
\mathcal{L}_{(8)}^{\mathrm{IH}}=-\frac{1}{4} \frac{\partial \mathcal{L}_{(4)}}{\partial F^{\alpha \beta}} \frac{\partial \mathcal{L}_{(6)}^{\mathrm{H}}}{\partial F_{\alpha \beta}}+\frac{1}{32} \frac{\partial^2 \mathcal{L}_{(4)}}{\partial F^{\alpha \beta} \partial F^{\rho \sigma}}\frac{\partial \mathcal{L}_{(4)}}{\partial F_{\alpha \beta}} \frac{\partial \mathcal{L}_{(4)}}{\partial F_{\rho \sigma}}\, .
\label{eq:partsol8}
\end{equation}
In this way, the theory is determined up to the eight-derivative level once the set of Lagrangians $\mathcal{L}_{(4)}$, $\mathcal{L}_{(6)}^{\mathrm{H}}$, $\mathcal{L}_{(8)}^{\mathrm{H}}$, is specified, so our final task consist in characterizing these. 

In order to characterize the Lagrangians that are invariant under a rotation of $F$ and $\star F$, it is useful to introduce the vector of 2-forms
\begin{equation}
\mathcal{F}^A=\begin{pmatrix}
		F\\
		\star F
	\end{pmatrix}\, ,
\end{equation}
where $A$ is an $\mathrm{SO}(2)$ index. Then, the only way of obtaining $\mathrm{SO}(2)$ invariant quantities is by considering the contraction $\mathcal{F}^A_{\mu\nu}\mathcal{F}^B_{\alpha\beta}\delta_{AB}$. Note that the contraction with the symplectic matrix $\sigma_{AB}$ also yields an invariant, but however it is not independent since  $\mathcal{F}^A_{\mu\nu}\mathcal{F}^B_{\alpha\beta}\sigma_{AB}=\mathcal{F}^A_{\mu\nu}\star\mathcal{F}^B_{\alpha\beta}\delta_{AB}$, and thus it has the same effect as applying the Hodge dual on the indices $\alpha\beta$.  Let us then evaluate this contraction explicitly,

\begin{align}
\mathcal{F}^A_{\mu\nu}\mathcal{F}^{B\, \alpha\beta}\delta_{AB}=F_{\mu\nu}F^{\alpha\beta}+\star F_{\mu\nu}\star F^{\alpha\beta}=F_{\mu\nu}F^{\alpha\beta}-6 F^{\rho\sigma}F_{[\rho\sigma}\tensor{\delta}{^{\alpha}_{\mu}}\tensor{\delta}{^{\beta}_{\nu]}}=4\tensor{T}{^{[\alpha}_{[\mu}}\tensor{\delta}{^{\beta]}_{\nu]}}\, ,
\end{align}
where $\tensor{T}{^{\alpha}_{\mu}}$ is the Maxwell stress tensor as defined in \req{eq:MaxTmunu}. Thus, this result tells us that any $\mathrm{SO}(2)$ invariant quantity must depend on $F_{\mu\nu}$ only through the stress tensor $T_{\mu\nu}$. Therefore, we conclude that the homogeneous part of the Lagrangians $\mathcal{L}_{(n)}^{\rm H}$ (including $\mathcal{L}_{(4)}=\mathcal{L}_{(4)}^{\rm H}$) must be of the form
\begin{equation}
\mathcal{L}_{(n)}^{\rm H}=\mathcal{L}_{(n)}^{\rm H}\left (g^{\mu\nu}, T_{\mu\nu}, R_{\mu\nu\rho\sigma},\nabla_{\alpha}R_{\mu\nu\rho\sigma},\ldots\right)\, .
\label{eq:homparteqs}
\end{equation}
This result together with the relations \req{eq:l6inh} and \req{eq:partsol8} completely characterizes the set of theories that have a duality-invariant constitutive relation \req{RotCons} to the eight-derivative level.

\subsection{Invariance of Einstein's equations} \label{subsec:inveineq}
We have just obtained the conditions the Lagrangian must satisfy in order for the constitutive relation \req{RotCons} between $H$ and $F$ to be invariant under a duality rotation. Now the question is whether these conditions ensure the invariance of Einstein's equations as well. For that, we see first in equation \req{RotEE} that the energy-momentum tensor $\hat T_{\mu\nu}$ (see \req{eq:hatTmunu}) is exactly invariant under a rotation, so we just have to make sure that the quantity $\mathcal{E}_{\mu\nu}$ --- defined in \eqref{eq:Emunu1} --- remains invariant. At this point it is convenient to study first those theories which are algebraic in the curvature tensor, since the proof for generic theories with covariant derivatives of the curvature is a direct generalization of that.

For algebraic theories, the tensor $\mathcal{E}_{\mu\nu}$ \eqref{eq:Emunu1} can in general be written as \cite{Cano:2020qhy}

\begin{equation}\label{EmunuEinst}
\mathcal{E}_{\mu\nu}=-\tensor{P}{_{(\mu}^{\rho \sigma \gamma}} R_{\nu) \rho \sigma \gamma}-2 \nabla^\sigma \nabla^\rho P_{(\mu| \sigma | \nu)\rho} +\frac{1}{2}g_{\mu \nu} \left(\mathcal{L}-\frac{1}{2}F_{\alpha\beta}\frac{\partial \mathcal{L}}{\partial F_{\alpha\beta}}\right)\, ,
\end{equation}
where 

\begin{equation}
P_{\mu\nu\rho\sigma}=\frac{\partial \ele}{\partial R^{\mu\nu\rho\sigma}}\, .
\end{equation}

\noindent
Let us then expand this in powers of $\ell$,

\begin{equation}
	\mathcal{E}_{\mu\nu}=\ell^2	\mathcal{E}^{(4)}_{\mu\nu}+\ell^4\mathcal{E}^{(6)}_{\mu\nu}+\ell^6 \mathcal{E}^{(8)}_{\mu\nu} \ldots\, ,
\end{equation}
where every term $\mathcal{E}^{(n)}_{\mu\nu}$ is computed from the corresponding Lagrangian $\mathcal{L}_{(n)}$. Let us then examine these terms. Notice that the Lagrangian $\mathcal{L}_{(4)}$ only depends on $F_{\mu\nu}$ through the Maxwell stress-energy tensor. Since every monomial $\mathcal{L}_{i}$ in the Lagrangian satisfies that $F_{\alpha\beta}\frac{\partial \mathcal{L}_i}{\partial F_{\alpha\beta}}\propto \mathcal{L}_{i}$, by looking at \req{EmunuEinst} we conclude that the tensor $\mathcal{E}^{(4)}_{\mu\nu}$ also depends on $F_{\mu\nu}$ through $T_{\mu\nu}$ only. We can express this fact by writing $\mathcal{E}^{(4)}_{\mu\nu}=\mathcal{E}^{(4)}_{\mu\nu}(T)$. Now, since under a duality transformation $T_{\mu\nu}$ is invariant up to terms of order $\ell^2$, we already conclude that the Einstein's equations are invariant up to terms of order $\ell^4$. 

Let us now see what happens with the $\mathcal{O}(\ell^4)$ and $\mathcal{O}(\ell^6)$ terms. First, we remind that we can split $\mathcal{L}_{(6)}$ and $\mathcal{L}_{(8)}$ in a homogeneous plus an inhomogeneous part, $\mathcal{L}_{(6)}=\mathcal{L}_{(6)}^{\rm H}+\mathcal{L}_{(6)}^{\rm IH}$ and $\mathcal{L}_{(8)}=\mathcal{L}_{(8)}^{\rm H}+\mathcal{L}_{(8)}^{\rm IH}$, where 
\begin{equation}
\mathcal{L}_{(6)}^{\rm IH}=-\frac{1}{8}\frac{\partial \mathcal{L}_{(4)}}{\partial \tensor{F}{^{\alpha\beta}}}\frac{\partial \mathcal{L}_{(4)}}{\partial \tensor{F}{_{\alpha\beta}}}\, , \qquad \mathcal{L}_{(8)}^{\rm IH}=-\frac{1}{4} \frac{\partial \mathcal{L}_{(6)}^{\rm H}}{\partial F^{\alpha \beta}} \frac{\partial \mathcal{L}_{(4)}}{\partial F_{\alpha \beta}}+\frac{1}{32} \frac{\partial^2 \mathcal{L}_{(4)}}{\partial F^{\alpha \beta} F^{\rho \sigma} } \frac{\partial \mathcal{L}}{\partial F_{\alpha \beta}} \frac{\partial \mathcal{L}}{\partial F_{\rho \sigma}}\,.
\end{equation}
Correspondingly, each $\mathcal{E}^{(n)}_{\mu\nu}$ splits as
\begin{equation}
\label{eq:inhomparts}
\mathcal{E}^{(6)}_{\mu\nu}=\mathcal{E}^{\rm H(6)}_{\mu\nu}+\mathcal{E}^{\rm IH (6)}_{\mu\nu}\, , \qquad  \mathcal{E}^{(8)}_{\mu\nu}=\mathcal{E}^{\rm H(8)}_{\mu\nu}+\mathcal{E}^{\rm IH (8)}_{\mu\nu}\,.
\end{equation}
Now it is useful to rewrite the terms coming from the homogeneous parts, $\mathcal{E}^{{\rm H}\, (n)}_{\mu\nu}$, in terms of the exactly-invariant tensor $\hat T_{\mu\nu}$. We recall that it is related to the Maxwell stress-energy tensor $T_{\mu\nu}$ by
\begin{equation}
	T_{\mu\nu}=\hat  T_{\mu\nu}+\frac{1}{2}F_{\langle \mu|\alpha}\frac{\partial \mathcal{L}}{\partial \tensor{F}{^{|\nu\rangle}_{\alpha}}}\, .
\end{equation}
Since $\mathcal{L}_{(6)}^{\rm H}$ and $\mathcal{L}_{(8)}^{\rm H}$ are built only out of the Maxwell stress-energy tensor $T_{\mu\nu}$, for the same reason as before it follows that $\mathcal{E}^{\rm H(6)}_{\mu\nu}$ and  $\mathcal{E}^{\rm H(8)}_{\mu\nu}$ depend on $F_{\mu\nu}$ only through $T_{\mu\nu}$. Hence we have
\begin{eqnarray}
\nonumber
\mathcal{E}^{(4)}_{\mu\nu}(T)&=&\mathcal{E}^{(4)}_{\mu\nu}(\hat T+\delta T)=\mathcal{E}^{(4)}_{\mu\nu}(\hat T)+\frac{\ell^2}{2}\frac{\delta \mathcal{E}^{(4)}_{\mu\nu}}{\delta T_{\alpha\beta}} (\hat{T}) \circ F_{\langle \alpha|\sigma}\frac{\partial \mathcal{L}_{(4)}}{\partial \tensor{F}{^{|\beta\rangle}_{\sigma}}}+\frac{\ell^4}{2}\frac{\delta \mathcal{E}^{(4)}_{\mu\nu}}{\delta T_{\alpha\beta}} (\hat{T}) \circ F_{\langle \alpha|\sigma}\frac{\partial \mathcal{L}_{(6)}}{\partial \tensor{F}{^{|\beta\rangle}_{\sigma}}}\\ \label{eq:expane4} &+& \frac{\ell^4}{8}\frac{\delta^2 \mathcal{E}^{(4)}_{\mu\nu}}{\delta T_{\alpha\beta} \delta T_{\rho \sigma}}(\hat{T}) \circ  F_{\langle \alpha|\lambda}\frac{\partial \mathcal{L}_{(4)}}{\partial \tensor{F}{^{|\beta\rangle}_{\lambda}}} \circ F_{\langle \rho|\gamma}\frac{\partial \mathcal{L}_{(4)}}{\partial \tensor{F}{^{|\sigma\rangle}_{\gamma}}}+\mathcal{O}(\ell^6)\,,  \\ \label{eq:expane6}
\mathcal{E}^{\rm H(6)}_{\mu\nu}(T)&=&\mathcal{E}^{\rm H(6)}_{\mu\nu}(\hat{T})+\frac{\ell^2}{2}\frac{\delta \mathcal{E}^{\rm H(6)}_{\mu\nu}}{\delta T_{\alpha\beta}} (\hat{T}) \circ F_{\langle \alpha|\sigma}\frac{\partial \mathcal{L}_{(4)}}{\partial \tensor{F}{^{|\beta\rangle}_{\sigma}}}+\mathcal{O}(\ell^4)\,,\\ \label{eq:expane8}
\mathcal{E}^{\rm H(8)}_{\mu\nu}(T)&=&\mathcal{E}^{\rm H(8)}_{\mu\nu}(\hat{T})+\mathcal{O}(\ell^2)\,,
\end{eqnarray}
where we have defined:
\begin{equation}
\begin{split}
\frac{\delta  }{\delta \mathcal{B}_{\mu_1 \dots \mu_p}} \circ \,  \mathcal{C}_{\mu_1 \dots \mu_p}=&\frac{\partial  }{\partial \mathcal{B}_{\mu_1 \dots \mu_p}} \mathcal{C}_{\mu_1 \dots \mu_p}+\frac{\partial  }{\partial \nabla_\nu \mathcal{B}_{\mu_1 \dots \mu_p}} \nabla_\nu\mathcal{C}_{\mu_1 \dots \mu_p}\\&+\frac{\partial  }{\partial \nabla_{\nu_{1}} \nabla_{\nu_2} \mathcal{B}_{\mu_1 \dots \mu_p}}\nabla_{\nu_{1}} \nabla_{\nu_2}\mathcal{C}_{\mu_1 \dots \mu_p}+\ldots\, ,
\end{split}
\end{equation}
being $\mathcal{B}$ and $\mathcal{C}$ arbitrary tensors. Taking into account that $\hat{T}_{\mu\nu}=T_{\mu\nu}+\mathcal{O}(\ell^2)$, we can express \eqref{eq:expane4}, \eqref{eq:expane6} and \eqref{eq:expane8} in the form:
\begin{eqnarray}
\nonumber
\mathcal{E}^{(4)}_{\mu\nu}(T)&=&\mathcal{E}^{(4)}_{\mu\nu}(\hat T)+\frac{\ell^2}{2}\frac{\delta \mathcal{E}^{(4)}_{\mu\nu}}{\delta T_{\alpha\beta}} (\hat{T}) \circ F_{\langle \alpha|\sigma}\frac{\partial \mathcal{L}_{(4)}}{\partial \tensor{F}{^{|\beta\rangle}_{\sigma}}}+\frac{\ell^4}{2}\frac{\delta \mathcal{E}^{(4)}_{\mu\nu}}{\delta T_{\alpha\beta}} (T) \circ F_{\langle \alpha|\sigma}\frac{\partial \mathcal{L}_{(6)}}{\partial \tensor{F}{^{|\beta\rangle}_{\sigma}}}\\ \label{eq:expane4mejor} &+& \frac{\ell^4}{8}\frac{\delta^2 \mathcal{E}^{(4)}_{\mu\nu}}{\delta T_{\alpha\beta} \delta T_{\rho \sigma}}(T) \circ  F_{\langle \alpha|\lambda}\frac{\partial \mathcal{L}_{(4)}}{\partial \tensor{F}{^{|\beta\rangle}_{\lambda}}} \circ F_{\langle \rho|\gamma}\frac{\partial \mathcal{L}_{(4)}}{\partial \tensor{F}{^{|\sigma\rangle}_{\gamma}}}+\mathcal{O}(\ell^6)\,,  \\
\mathcal{E}^{\rm H(6)}_{\mu\nu}(T)&=&\mathcal{E}^{\rm H(6)}_{\mu\nu}(\hat{T})+\frac{\ell^2}{2}\frac{\delta \mathcal{E}^{\rm H(6)}_{\mu\nu}}{\delta T_{\alpha\beta}} (T) \circ F_{\langle \alpha|\sigma}\frac{\partial \mathcal{L}_{(4)}}{\partial \tensor{F}{^{|\beta\rangle}_{\sigma}}}+\mathcal{O}(\ell^4)\,,\\
\mathcal{E}^{\rm H(8)}_{\mu\nu}(T)&=&\mathcal{E}^{\rm H(8)}_{\mu\nu}(\hat{T})+\mathcal{O}(\ell^2)\,,
\end{eqnarray}

\noindent
Let us further rewrite \eqref{eq:expane4mejor} as follows,
\begin{eqnarray}
\nonumber
\mathcal{E}^{(4)}_{\mu\nu}(T)&=&\mathcal{E}^{(4)}_{\mu\nu}(\hat T)+\frac{\ell^2}{2}\frac{\delta \mathcal{E}^{(4)}_{\mu\nu}}{\delta T_{\alpha\beta}} (T) \circ F_{\langle \alpha|\sigma}\frac{\partial \mathcal{L}_{(4)}}{\partial \tensor{F}{^{|\beta\rangle}_{\sigma}}}+\frac{\ell^4}{2}\frac{\delta \mathcal{E}^{(4)}_{\mu\nu}}{\delta T_{\alpha\beta}} (T) \circ F_{\langle \alpha|\sigma}\frac{\partial \mathcal{L}_{(6)}}{\partial \tensor{F}{^{|\beta\rangle}_{\sigma}}}\\ &-& \frac{\ell^4}{8}\frac{\delta^2 \mathcal{E}^{(4)}_{\mu\nu}}{\delta T_{\alpha\beta} \delta T_{\rho \sigma}}(T) \circ  F_{\langle \alpha|\lambda}\frac{\partial \mathcal{L}_{(4)}}{\partial \tensor{F}{^{|\beta\rangle}_{\lambda}}} \circ F_{\langle \rho|\gamma}\frac{\partial \mathcal{L}_{(4)}}{\partial \tensor{F}{^{|\sigma\rangle}_{\gamma}}}+\mathcal{O}(\ell^6)\,,
\end{eqnarray}

\noindent
where we have replaced $\hat T=T-\delta T$ in the $\mathcal{O}(\ell^2)$ term and expanded in $\delta T$ one again. 
Now, taking into account the following identity,

\begin{equation}
	\frac{\delta }{\delta T_{\alpha\beta}}\circ F_{\langle \alpha|\sigma}\mathcal{A}_{|\beta\rangle}{}^{\sigma}=	\frac{1}{2}\frac{\delta }{\delta F_{\alpha\beta}} \circ \mathcal{A}_{\alpha \beta}\,,
\end{equation}
which is valid for any antisymmetric tensor $\mathcal{A}_{\alpha \beta}$, we find that

\begin{equation}
\label{eq:emunuexpanmejor}
\begin{split}
\mathcal{E}_{\mu\nu}&=\ell^2\mathcal{E}^{(4)}_{\mu\nu}(\hat T)+\ell^4\left(\mathcal{E}^{\rm H(6)}_{\mu\nu} (\hat{T})+\mathcal{E}^{\rm IH(6)}_{\mu\nu}(T)+\frac{1}{4}\frac{\delta \mathcal{E}^{(4)}_{\mu\nu} }{\delta F_{\alpha\beta}}(T)\circ \frac{\partial \mathcal{L}_{(4)}}{\partial \tensor{F}{^{\alpha\beta}}}\right)+ \ell^6 \Bigg ( \mathcal{E}^{\rm H(8)}_{\mu\nu} (\hat{T})+\mathcal{E}^{\rm IH(8)}_{\mu\nu}(T) \\& \left.+\frac{1}{4}\frac{\delta \mathcal{E}^{(4)}_{\mu\nu}}{\delta F_{\alpha\beta}} (T) \circ \frac{\partial \mathcal{L}_{(6)}}{\partial F_{\alpha \beta}}-\frac{1}{16}\frac{\delta^2 \mathcal{E}^{(4)}_{\mu\nu}}{\delta F_{\alpha\beta} \delta F_{\rho \sigma}}(T) \circ  \frac{\partial \mathcal{L}_{(4)}}{\partial F^{\alpha \beta} } \circ \frac{\partial \mathcal{L}_{(4)}}{\partial F^{\rho \sigma} }+\frac{1}{4}\frac{\delta \mathcal{E}^{\rm H(6)}_{\mu\nu}}{\delta F_{\alpha\beta}} (T) \circ \frac{\partial \mathcal{L}_{(4)}}{\partial F^{\alpha \beta}}\right ) +\mathcal{O}(\ell^8)\,,
\end{split}
\end{equation}

\noindent
Finally, one can prove the following identities:
\begin{eqnarray}
\label{eq:firstidentityeins}
& &\mathcal{E}^{\rm IH (6)}_{\mu\nu}(T)+	\frac{1}{4}\frac{\delta \mathcal{E}^{(4)}_{\mu\nu} }{\delta F_{\alpha\beta}}(T)\circ \frac{\partial \mathcal{L}_{(4)}}{\partial \tensor{F}{^{\alpha\beta}}}=0\, ,\\ \nonumber
 & & \mathcal{E}^{\rm IH(8)}_{\mu\nu}(T) +\frac{1}{4}\frac{\delta \mathcal{E}^{(4)}_{\mu\nu}}{\delta F_{\alpha\beta}} (T) \circ \frac{\partial \mathcal{L}_{(6)}}{\partial F_{\alpha \beta}}-\frac{1}{16}\frac{\delta^2 \mathcal{E}^{(4)}_{\mu\nu}}{\delta F_{\alpha\beta} \delta F_{\rho \sigma}}(T) \circ  \frac{\partial \mathcal{L}_{(4)}}{\partial F^{\alpha \beta} } \circ \frac{\partial \mathcal{L}_{(4)}}{\partial F^{\rho \sigma} }\\& &\label{eq:secondidentityeins}+\frac{1}{4}\frac{\delta \mathcal{E}^{\rm H(6)}_{\mu\nu}}{\delta F_{\alpha\beta}} (T) \circ \frac{\partial \mathcal{L}_{(4)}}{\partial F^{\alpha \beta}}=0\,.
\end{eqnarray}

\noindent
In order to show this, it is useful to take into account the following property,

\begin{equation}
\label{eq:propnabladerfunc}
\frac{\delta \nabla_{\mu_{1}} \nabla_{\mu_2} \mathcal{Q}_{\nu_1 \dots \nu_p}}{\delta F_{\alpha \beta}}\circ \mathcal{A}_{\alpha \beta}=\nabla_{\mu_1}  \nabla_{\mu_2} \left ( \frac{\partial \mathcal{Q}_{\nu_1 \dots \nu_p}}{\partial F_{\alpha \beta}} \mathcal{A}_{\alpha \beta} \right ) \,,
\end{equation}


\noindent
where $\mathcal{Q}_{\nu_1 \dots \nu_p}$ is an arbitrary tensor which depends algebraically on $F_{\mu \nu}$ and where $\mathcal{A}_{\alpha \beta}$ is any antisymmetric tensor.  Let us illustrate how to use \eqref{eq:propnabladerfunc} to prove the first identity \eqref{eq:firstidentityeins}. On the one hand, since $\mathcal{L}_{(4)}$ depends algebraically on the curvature, we have, explicitly,

\begin{equation}
\begin{split}
\frac{\delta \mathcal{E}^{(4)}_{\mu\nu} }{\delta F_{\alpha\beta}}\circ \frac{\partial \mathcal{L}_{(4)}}{\partial \tensor{F}{^{\alpha\beta}}}&=-\hat{P}^{(4)}_{(\mu}{}^{\rho \sigma \gamma} R_{\nu) \rho \sigma \gamma} +\frac{1}{2}g_{\mu \nu} \frac{\partial \mathcal{L}_{(4)}}{\partial \tensor{F}{^{\alpha\beta}}} \frac{\partial }{\partial F_{\alpha\beta}}\left(\mathcal{L}_{(4)}-\frac{1}{2}F_{\rho\sigma}\frac{\partial \mathcal{L}_{(4)}}{\partial F_{\rho\sigma}}\right)\\& -2 \frac{\delta \nabla^\sigma \nabla^\rho P^{(4)}_{(\mu| \sigma | \nu)\rho}}{ \delta F_{\alpha \beta}} \circ\frac{\partial \mathcal{L}_{(4)}}{\partial F^{\alpha\beta}}\\&=-\hat{P}^{(4)}_{(\mu}{}^{\rho \sigma \gamma} R_{\nu) \rho \sigma \gamma} -2\nabla^\sigma \nabla^\rho \hat{P}^{(4)}_{(\mu| \sigma | \nu)\rho} +\frac{1}{2}g_{\mu \nu} \frac{\partial \mathcal{L}_{(4)}}{\partial \tensor{F}{^{\alpha\beta}}} \frac{\partial }{\partial F_{\alpha\beta}}\left(\mathcal{L}_{(4)}-\frac{1}{2}F_{\rho\sigma}\frac{\partial \mathcal{L}_{(4)}}{\partial F_{\rho\sigma}}\right)\,,\\
\end{split}
\label{eq:inveinsl4}
\end{equation}

\noindent
where 
\begin{equation}
P^{(4)}{}^{\mu \nu \rho \sigma}=\frac{\partial \mathcal{L}_{(4)}}{\partial R_{\mu \nu \rho \sigma}}\, , \quad \hat{P}^{(4)}{}^{\mu \nu \rho \sigma}=\frac{\partial \mathcal{L}_{(4)}}{\partial \tensor{F}{^{\alpha\beta}}}\frac{\partial P^{(4)}{}^{\mu \nu \rho \sigma}}{\partial \tensor{F}{_{\alpha\beta}}}\, .
\end{equation}
After some direct computations, we recognize at the last line of \eqref{eq:inveinsl4} the term $-4\mathcal{E}^{\rm IH (6)}_{\mu\nu}(T)$, so we prove \eqref{eq:firstidentityeins}. Showing that \eqref{eq:secondidentityeins} holds is a more intricate task, but it can be done by using \eqref{eq:propnabladerfunc} and following a completely analogous procedure to that of the proof of \eqref{eq:firstidentityeins}.
Consequently, we have shown that the Einstein's equations can be written as

\begin{equation}
	\mathcal{E}_{\mu\nu}=\ell^2\mathcal{E}^{(4)}_{\mu\nu}(\hat T)+\ell^4\mathcal{E}^{\rm H(6)}_{\mu\nu}(\hat T)+\ell^6\mathcal{E}^{\rm H(8)}_{\mu\nu}(\hat T)+\mathcal{O}(\ell^8)\, ,
\end{equation}
so that they are manifestly duality invariant to order $\ell^6$. Thus, the conclusion is that the invariance of the constitutive relation implies the invariance of Einstein's equations, so the conditions found in the previous subsection are necessary and sufficient in order for a theory with algebraic dependence on the Riemann tensor to be duality invariant. 

These results can be generalized to the case of theories that contain covariant derivatives of the curvature. The proof of the invariance of Einstein's equations in that case can be obtained along similar lines, although it is slightly more technical and thus we include it in Appendix~\ref{app:inv}.

\section{All duality-invariant theories up to eight derivatives}\label{sec:all8d}

The goal of this section is to obtain explicitly the Lagrangian of the most general duality-invariant theory up to eight-derivative terms. For that, we are going to exploit the equations \eqref{eq:l6inh}, \eqref{eq:partsol8} and \eqref{eq:homparteqs}. In addition, we will also assume that parity is preserved, so that we will discard parity-breaking operators. 

We start by analyzing the fourth-derivative terms. Since all the dependence of $\mathcal{L}_{(4)}$ on $F_{\mu \nu}$ must be through the Maxwell stress-energy tensor $T_{\mu \nu}$, we observe that the most general duality-invariant four-derivative Lagrangian takes the form
\begin{equation}
 \mathcal{L}_{(4)}=\alpha_1 T_{\mu \nu} T^{\mu \nu}+\alpha_2 R^{\mu \nu} T_{\mu \nu}+ \alpha_3 \mathcal{X}_4+ \alpha_4 R^{\mu \nu} R_{\mu \nu}+ \alpha_5 R^2,
\label{eq:l4def}
\end{equation}
where $\mathcal{X}_4$ denotes the topological Gauss-Bonnet density, given by $\mathcal{X}_4=R_{\mu\nu\rho\sigma}R^{\mu\nu\rho\sigma}-4R_{\mu\nu}R^{\mu\nu}+R^2$. Now we move into the most general six-derivative Lagrangian $\mathcal{L}_{(6)}$. In the previous section we decomposed it into a homogeneous part $\mathcal{L}_{(6)}^{\mathrm{H}}$ which takes the functional form given at \eqref{eq:homparteqs} and an inhomogeneous part $\mathcal{L}_{(6)}^{\mathrm{IH}}$ which is obtained from the four-derivative Lagrangian and whose particular form we rewrite here:
\begin{equation}
\mathcal{L}_{(6)}^{\mathrm{IH}}=-\frac{1}{8} \frac{\partial \mathcal{L}_{(4)}}{\partial F^{\alpha \beta}} \frac{\partial \mathcal{L}_{(4)}}{\partial F_{\alpha \beta}}\, .
\end{equation}
Now, taking into account that
\begin{equation}
\frac{\partial \mathcal{L}_{(4)}}{\partial F_{\rho \sigma}}=4 \alpha_1 T^{\mu [\rho} F_\mu{}^{\sigma]}+2 \alpha_2 \hat{R}^{\mu [\rho} F_\mu{}^{\sigma]}\, ,
\end{equation}
where $\hat{R}_{\mu \nu}=R_{\langle \mu \nu \rangle}$, we obtain the following expression for $\mathcal{L}_{(6)}^{\mathrm{IH}}$,
\begin{equation}
\begin{split}
\mathcal{L}_{(6)}^{\mathrm{IH}}&=-2 \alpha_1^2 T^{\mu [\rho} F_\mu{}^{\sigma]}  T_{\nu \rho} F^\nu{}_{\sigma}-2  \alpha_1 \alpha_2 T^{\mu [\rho} F_\mu{}^{\sigma]} \hat{R}_{\nu \rho} F^\nu{}_{\sigma}-\frac{1}{2} \alpha_2^2 \hat{R}^{\mu [\rho} F_\mu{}^{\sigma]} \hat{R}_{\nu \rho} F^\nu{}_{\sigma}\,.
\end{split}
\label{eq:particularl4}
\end{equation}

Once we have obtained the inhomogeneous part, we now write down the most general homogeneous Lagrangian $\mathcal{L}_{(6)}^\mathrm{H}$ that preserves duality and parity. This can be seen to have the following form:
\begin{equation}
\mathcal{L}_{(6)}^{\mathrm{H}}=\mathcal{R}_{(6)}^{\mu \nu \alpha \beta} T_{\mu \nu} T_{\alpha \beta} + \mathcal{R}_{(6)}^{\mu \nu} T_{\mu \nu}+ \mathcal{R}_{(6)}\, ,
\label{eq:homl6sol}
\end{equation}
where $\mathcal{R}_{(6)}^{\mu \nu \alpha \beta}$ and $\mathcal{R}_{(6)}^{\mu \nu}$ are tensors formed out of the curvature and $\mathcal{R}_{(6)}$ is the general six-derivative Lagrangian for the metric. After discarding trivial terms and total derivatives, the general form of these quantities reads \cite{Fulling:1992vm}
\begin{eqnarray}
\mathcal{R}_{(6)}^{\mu \nu \alpha \beta}&=& \beta_1 R^{\mu \alpha\nu \beta} +\beta_2 R^{\mu \alpha} g^{\beta \nu}+\beta_3 R g^{\mu \alpha} g^{\beta \nu}\, , \\ \notag
\mathcal{R}_{(6)}^{\mu \nu}&=&\beta_4 R_{\alpha \beta \gamma}{}^\mu R^{\alpha \beta \gamma \nu}+ \beta_5 R^{\rho \mu \alpha \nu} R_{\rho \alpha}+\beta_6 R^{\mu \alpha} R_{\alpha}{}^\nu+\beta_7 R R^{\mu \nu} \\&+& \beta_{8} \nabla^{(\mu} \nabla^{\nu)} R+\beta_{9} \nabla^2 R^{\mu \nu} \, , \\ \notag
\mathcal{R}_{(6)}&=&  \beta_{10} R^{ \rho \sigma \mu \nu} R_{\mu \nu}{}^{\lambda \eta} R_{ \lambda \eta \rho \sigma}+ \beta_{11}R R^{\mu \nu \rho \sigma} R_{\mu \nu \rho \sigma}+\beta_{12}R^{\mu \nu} R^{\rho \sigma} R_{\mu \rho \nu \sigma}+ \beta_{13}R^{\mu \nu} R_{\mu \alpha} R_{\nu}{}^\alpha\\&+&\beta_{14}R R^{\mu \nu} R_{\mu \nu}+\beta_{15} R^3+ \beta_{16} \nabla^\sigma R\nabla_\sigma R+ \beta_{17} \nabla_\sigma R_{\mu \nu} \nabla^\sigma R^{\mu \nu}\,.
\end{eqnarray}

The eight-derivative Lagrangian $\mathcal{L}_{(8)}$ also decomposes in a homogeneous part $\mathcal{L}_{(8)}^{\mathrm{H}}$ and an inhomogeneous one $\mathcal{L}_{(8)}^{\mathrm{IH}}$. This last piece was seen to be expressed in terms of the lower-derivative Lagrangians $\mathcal{L}_{(4)}$ and $\mathcal{L}_{(6)}^\mathrm{H}$ as
\begin{equation}
\mathcal{L}_{(8)}^{\mathrm{IH}}=-\frac{1}{4} \frac{\partial \mathcal{L}_{(4)}}{\partial F^{\alpha \beta}} \frac{\partial \mathcal{L}_{(6)}^{\mathrm{H}}}{\partial F_{\alpha \beta}}+\frac{1}{32} \frac{\partial^2 \mathcal{L}_{(4)}}{\partial F^{\alpha \beta} \partial F^{\rho \sigma}}\frac{\partial \mathcal{L}_{(4)}}{\partial F_{\alpha \beta}} \frac{\partial \mathcal{L}_{(4)}}{\partial F_{\rho \sigma}}\, .
\end{equation}
Taking into account that
\begin{eqnarray}
\frac{\partial \mathcal{L}_{(6)}^\mathrm{H}}{\partial F_{\rho \sigma}}&=& 4 T_{\alpha \beta} \hat{\mathcal{R}}_{(6)}^{\alpha \beta \nu [\rho} F_{\nu}{}^{\sigma]}+2\hat{\mathcal{R}}_{(6)}^{\nu[\rho} F_\nu{}^{\sigma]}\, , \\ \nonumber
\frac{\partial^2 \mathcal{L}_{(4)}}{\partial F_{\alpha \beta} F^{\rho \sigma} }&=& 4 \alpha_1 F^\alpha{}_{[\rho} F^\beta{}_{\sigma]} -2 \alpha_1 F_{\rho \sigma} F^{\alpha \beta} +8 \alpha_1 T_{[\rho}{}^{[\alpha} \delta_{\sigma]}{}^{\beta]}+\alpha_1 F^2 \delta_{[\rho}{}^{[\alpha} \delta_{\sigma]}{}^{\beta]}\\ &+& 2 \alpha_2 \hat{R}_{[\rho}{}^{[\alpha} \delta_{\sigma]}{}^{\beta]}\, ,
\end{eqnarray}
where  a hat over $\mathcal{R}_{(6)}^{\alpha \beta \nu \rho}$ means that we take its traceless part over each pair of indices, we conclude that \eqref{eq:partsol8} is given by
\begin{equation}
\begin{split}
\mathcal{L}_{(8)}^{\mathrm{IH}}&=-4 \alpha_1 T_{\alpha \beta} \hat{\mathcal{R}}_{(6)}^{\alpha \beta \nu [\rho} F_{\nu}{}^{\sigma]} T_{\mu \rho} F^\mu{}_\sigma-2 \alpha_2 T_{\alpha \beta} \hat{\mathcal{R}}_{(6)}^{\alpha \beta \nu [\rho} F_{\nu}{}^{\sigma]} \hat{R}_{\mu \rho} F^\mu{}_\sigma-2\alpha_1 \hat{\mathcal{R}}_{(6)}^{\nu[\rho} F_\nu{}^{\sigma]} T_{\mu \rho} F^\mu{}_\sigma\\&-\alpha_2 \hat{\mathcal{R}}_{(6)}^{\nu[\rho} F_\nu{}^{\sigma]} \hat{R}_{\mu \rho} F^\mu{}_\sigma+2 \alpha_1^3 T_{\mu \alpha} F^\mu{}_\beta T^{\nu \rho} F_\nu{}^\sigma F^\alpha{}_{[\rho} F^\beta{}_{\sigma]}-\alpha_1^3 T_{\mu \alpha} F^\mu{}_\beta T^{\nu \rho} F_\nu{}^\sigma F_{\rho \sigma} F^{\alpha \beta}\\&+4 \alpha_1^3 T_{\mu \alpha} F^\mu{}_\beta T^{\nu \rho} F_\nu{}^\sigma T_{[\rho}{}^{[\alpha} \delta_{\sigma]}{}^{\beta]}+1/2 \alpha_1^3 F^2 T^{\mu [\rho} F_{\mu}{}^{\sigma]} T_{\nu \rho} F^{\nu}{}_{\sigma}\\&+2 \alpha_1^2 \alpha_2 T^{\nu \rho} F_{\nu}{}^\sigma \hat{R}_{\mu \alpha}F^\mu{}_\beta F^\alpha{}_{[\rho} F^\beta{}_{\sigma]}-\alpha_1^2 \alpha_2 T^{\nu \rho} F_{\nu}{}^\sigma \hat{R}_{\mu \alpha}F^\mu{}_\beta F_{\rho \sigma} F^{\alpha \beta}\\&+4 \alpha_1^2 \alpha_2 T^{\nu \rho} F_{\nu}{}^\sigma \hat{R}_{\mu \alpha}F^\mu{}_\beta T_{[\rho}{}^{[\alpha} \delta_{\sigma]}{}^{\beta]}+ \alpha_1^2 \alpha_2 T_{\mu \alpha} F^\mu{}_\beta T^{\nu \rho} F_\nu{}^\sigma \hat{R}_{[\rho}^{[\alpha}\delta_{\sigma]}{}^{\beta]}\\&+1/2 \alpha_1^2 \alpha_2 F^2 T^{\mu [\rho} F_\mu{}^{\sigma]} \hat{R}_{\nu \rho} F^\nu{}_\sigma+ 1/2\alpha_1 \alpha_2^2 \hat{R}_{\mu \alpha} F^\mu{}_\beta \hat{R}^{\nu \rho} F_\nu{}^\sigma F^\alpha{}_{[\rho} F^\beta{}_{\sigma]}\\&-1/4\alpha_1 \alpha_2^2 \hat{R}_{\mu \alpha} F^\mu{}_\beta \hat{R}^{\nu \rho} F_\nu{}^\sigma F_{\rho \sigma} F^{\alpha \beta}+ \alpha_1 \alpha_2^2 \hat{R}_{\mu \alpha} F^\mu{}_\beta \hat{R}^{\nu \rho} F_\nu{}^\sigma T_{[\rho}{}^{[\alpha} \delta_{\sigma]}{}^{\beta]}\\&+ \alpha_1 \alpha_2^2 T^{\nu \rho} F_{\nu}{}^\sigma \hat{R}_{\mu \alpha}F^\mu{}_\beta  \hat{R}_{[\rho}^{[\alpha}\delta_{\sigma]}{}^{\beta]}+1/8 \alpha_1 \alpha_2^2 F^2 \hat{R}^{\mu [\rho} F_\mu{}^{\sigma]} \hat{R}_{\nu \rho} F^\nu{}_\sigma\\&+1/4 \alpha_2^3 \hat{R}_{\mu \alpha} F^\mu{}_\beta \hat{R}^{\nu \rho} F_\nu{}^\sigma \hat{R}_{[\rho}^{[\alpha} \delta_{\sigma]}{}^{\beta]}\,.
\end{split}
\end{equation}

\noindent
Finally, the homogeneous part $\mathcal{L}_{(8)}^{\mathrm{H}}$ must take the form \eqref{eq:homparteqs} and hence it can be written as
\begin{equation}
\mathcal{L}_{(8)}^{\mathrm{H}}=\gamma_1(T_{\mu\nu}T^{\mu\nu})^2+\gamma_2\tensor{T}{_{\mu}^{\nu}}\tensor{T}{_{\nu}^{\alpha}}\tensor{T}{_{\alpha}^{\beta}}\tensor{T}{_{\beta}^{\mu}}+\mathcal{R}_{(8)}^{\mu \nu  \rho \sigma \alpha \beta} T_{\mu \nu} T_{\rho \sigma} T_{\alpha \beta} + \mathcal{R}_{(8)}^{\mu \nu \rho \sigma } T_{\mu \nu} T_{\rho \sigma }+ \mathcal{R}_{(8)}^{\mu \nu } T_{\mu \nu}+ \mathcal{R}_{(8)}\, .
\end{equation}
A list with all the possible tensors appearing in these expressions can be obtained from \cite{Fulling:1992vm}. We get the following general expressions,
\begin{eqnarray}
\mathcal{R}_{(8)}{}_{\mu \nu \rho \sigma \alpha \beta}&=& \gamma_3 R_{\mu \rho \nu \alpha} g_{\sigma \beta}+\gamma_4 R_{\mu \nu} g_{\rho \alpha} g_{\sigma \beta}+  \gamma_5 R_{\mu \rho} g_{\alpha \nu} g_{\sigma \beta}\, , \\ \notag
\mathcal{R}_{(8)}{}_{\mu \nu \rho \sigma} &=&\gamma_6 R_{\mu \nu} R_{\rho \sigma} +\gamma_{7} R_{\mu \rho} R_{\nu \sigma}+\gamma_{8} R R^{\mu \rho \nu \sigma}+\gamma_{9} R^{\alpha \beta}{}_{\mu \rho} R_{\alpha \beta \nu \sigma}+\gamma_{10} R^{\alpha}{}_{\mu}{}^{\beta}{}_{\nu} R_{\alpha \beta \rho \sigma}\\ \notag &+& \gamma_{11} R^{\alpha}{}_{\mu}{}^{\beta}{}_\rho R_{\alpha \nu \beta \sigma}+ \gamma_{12} \nabla_\mu \nabla_\rho R g_{\nu \sigma}+\gamma_{13} \nabla^2 R_{\mu \rho} g_{\nu \sigma}+\gamma_{14} R R_{\mu \rho} g_{\nu \sigma}+ \gamma_{15} R^{\alpha}{}_\mu R_{\alpha \rho} g_{\nu \sigma}\\ \notag &+& \gamma_{16}  R^{\alpha \beta} R_{\alpha \mu \beta \rho} g_{\nu \sigma}+\gamma_{17} R^{\alpha \beta \gamma}{}_\mu R_{\alpha \beta \gamma \rho} g_{\nu \sigma}+ \gamma_{18} \nabla^2 R g_{\mu \rho} g_{\nu \sigma}+\gamma_{19} R^2g_{\mu \rho} g_{\nu \sigma}\\&+& \gamma_{20}   R^{\alpha \beta} R_{\alpha \beta} g_{\mu \rho} g_{\nu \sigma}+\gamma_{21} R^{\alpha \beta \lambda \eta} R_{\alpha \beta \lambda \eta} g_{\mu \rho} g_{\nu \sigma}\, ,\\ \notag
\mathcal{R}_{(8)}{}_{\mu \nu} &=&\gamma_{22} \nabla_\mu \nabla_\nu \nabla^2 R
+\gamma_{23} \nabla^2 \nabla^2 R_{\mu \nu}+\gamma_{24} R \nabla_{\mu}\nabla_{\nu} R+\gamma_{25} R_{\mu \nu} \nabla^2 R +\gamma_{26} R^{\alpha \beta} \nabla_{\alpha} \nabla_{\beta} R_{\mu \nu} \\ \notag &+&\gamma_{27} R \nabla^2 R_{\mu \nu}+\gamma_{28} R^{\alpha \beta} \nabla_\mu \nabla_\nu R_{\alpha \beta}+\gamma_{29} R^{\alpha \beta} \nabla_\nu \nabla_\beta R_{\mu \alpha}+\gamma_{30} \nabla^{\alpha} \nabla_\mu R R_{\alpha \nu}\\ \notag &+& \gamma_{31} \nabla^{\rho} \nabla^\sigma R R_{\rho \mu \sigma \nu} +\gamma_{32} \nabla^2 R^{\rho \sigma} R_{\rho \mu \sigma \nu}+\gamma_{33} \nabla^\rho \nabla_\mu R^{\alpha \beta} R_{\alpha \rho \beta \nu}+\gamma_{34}\nabla^\beta \nabla^\rho R^{\alpha}{}_\mu R_{\alpha \beta \rho \nu}  \\ \notag &+& \gamma_{35} R^{\alpha \beta \rho \sigma} \nabla_\mu \nabla_\nu R_{\alpha \beta \rho \sigma}+   \gamma_{36} \nabla_\mu R^{\alpha \beta} \nabla_\nu R_{\alpha \beta}+\gamma_{37} \nabla_\alpha R_{\beta \mu} \nabla^\alpha R^{\beta}{}_\nu+\gamma_{38} \nabla^\beta R^{\alpha}{}_\mu \nabla_\alpha R_{\beta \nu}\\ \notag &+&\gamma_{39} \nabla^\rho R_{\alpha \beta} \nabla_\nu R_{\alpha \rho \beta \mu}+ \gamma_{40} \nabla_\mu R \nabla_\nu R + \gamma_{41}\nabla_\alpha R \nabla_\nu R_\mu{}^\alpha+\gamma_{42}\nabla_\alpha R \nabla^\alpha  R_{\mu \nu}\\ \notag &+&\gamma_{43} \nabla_\mu R^{\alpha \beta} \nabla_\beta R_{\alpha \nu}+\gamma_{44} \nabla^\sigma R^{\alpha \beta} \nabla_\sigma R_{\alpha \mu \beta \nu}+\gamma_{45} \nabla_{\mu} R^{\alpha \beta \sigma \lambda} \nabla_\nu R_{\alpha \beta \sigma \lambda}\\ \notag &+&\gamma_{46} \nabla^\lambda R^{\alpha \beta \sigma}{}_\mu \nabla_\lambda R_{\alpha \beta \sigma \nu}+\gamma_{47} R^2 R_{\mu \nu}+\gamma_{48} R R^\alpha{}_\mu R_{\alpha \nu}+\gamma_{49} R_{\mu \nu} R^{\alpha \beta} R_{\alpha \beta} \\ \notag &+&\gamma_{50} R^{\alpha \beta} R_{\alpha \mu} R_{\beta \nu}+\gamma_{51} R R^{\alpha \beta} R_{\alpha \mu \beta \nu}+\gamma_{52} R_\alpha{}^\sigma R_{\beta \mu \sigma \nu} +\gamma_{53} R^{\alpha \beta} R^\sigma{}_\mu R_{\alpha \sigma \beta \nu}\\ \notag &+& \gamma_{54} R R^{\alpha \beta \sigma}{}_\mu R_{\alpha \beta \sigma \nu}+\gamma_{55} R_{\mu \nu} R^{\alpha \beta \lambda \sigma} R_{\alpha \beta \lambda \sigma} +\gamma_{56} R^\alpha{}_\mu R^{\beta \rho \sigma}{}_\alpha R_{\beta \rho \sigma \nu} \\ \notag &+&\gamma_{57} R^{\alpha \beta} R^{\rho \sigma}{}_{\alpha \mu} R_{\rho \sigma \beta \nu}+ \gamma_{58} R^{\alpha \beta} R_\alpha{}^\rho{}_\beta{}^\sigma R_{\rho \mu \sigma \nu} +\gamma_{59} R^{\alpha \beta} R^\rho{}_\alpha{}^\sigma{}_\mu R_{\rho \beta \sigma \nu}\\ \notag &+&\gamma_{60} R^{\alpha \beta \rho \sigma} R_{\alpha \beta}{}^\lambda{}_\mu R_{\rho \sigma \lambda \nu}+ \gamma_{61} R^{\alpha \beta \rho \sigma} R_{\alpha}{}^\lambda{}_{\rho \mu} R_{\beta \lambda \sigma \nu}+ \gamma_{62} R^{\alpha \beta \rho \sigma} R_{\alpha \beta \rho}{}^\lambda R_{\sigma \mu \lambda \nu}\,.
\end{eqnarray}
On the other hand, the list of the eight-derivative curvature invariants appearing in $\mathcal{R}_{(8)}$ can also be checked in \cite{Fulling:1992vm}. 

Let us note that here and also in the case of $\mathcal{L}_{(6)}^{\mathrm{H}}$ we are including a set of densities that spans the set of all duality-invariant Lagrangians, but they may be not linearly independent. Even though we are removing redundant terms in the $\mathcal{R}_{(n)}^{\mu_1\mu_2\ldots}$ tensors following \cite{Fulling:1992vm}, the densities formed by contracting these tensors with $T_{\mu\nu}$ may still be linearly dependent or be related up to total derivatives. The determination of a linearly independent set of generating densities for the Lagrangians $\mathcal{L}_{(6)}^{\mathrm{H}}$ and $\mathcal{L}_{(8)}^{\mathrm{H}}$ may be carried out elsewhere.

\section{Linear theories}\label{sec:linear}
As we have seen, any duality-invariant modification of Einstein-Maxwell theory requires the introduction of an infinite tower of higher-derivative terms, due to the non-linearity of this symmetry. Finding these terms becomes very hard and there seems to be no simple way of obtaining a formula for the $n$-th order density. Thus, here we focus on a subset of these theories that have a simpler form: those with linear Maxwell equations.

Let us consider the action
\begin{equation}\label{eq:linearth}
	S=\frac{1}{16\pi G}\int d^4x\sqrt{|g|}\big[R-\chi^{\mu\nu\rho\sigma}F_{\mu\nu}F_{\rho\sigma}\big]\, ,
\end{equation}
where $\chi^{\mu\nu\rho\sigma}$ is a tensor built out of the metric and the Riemann tensor. Without loss of generality, we can assume that it has the symmetries

\begin{equation}
\chi^{\mu\nu\rho\sigma}=-\chi^{\nu\mu\rho\sigma}=-\chi^{\mu\nu\sigma\rho}=\chi^{\rho\sigma\mu\nu}\, .
\end{equation}
Then, we are going to show that the equations of motion of this theory are invariant under duality rotations if and only if 

\begin{equation}\label{eq:chiinv}
	\tensor{(\star\chi)}{_{\mu\nu}^{\alpha\beta}}	\tensor{(\star\chi)}{_{\alpha\beta}^{\rho\sigma}}=-\tensor{\delta}{_{[\mu}^{[\rho}}\tensor{\delta}{_{\nu]}^{\sigma]}}\, ,
\end{equation}
where 

\begin{equation}
	\tensor{(\star\chi)}{_{\mu\nu}^{\alpha\beta}}=\frac{1}{2}\epsilon_{\mu\nu\lambda\gamma}\chi^{\lambda\gamma\alpha\beta}\, .
\end{equation}

Let us prove that \eqref{eq:chiinv} is indeed a necessary condition for duality invariance. For that, we start with the constitutive relation, which can be written as 
\begin{equation}\label{eq:constlinear0}
\star H_{\mu\nu}=-\tensor{\chi}{_{\mu\nu}^{\alpha\beta}} F_{\alpha\beta}\, .
\end{equation}
If we consider the $\mathrm{SO}(2)$ transformation which sends $H \rightarrow -F$ and $F \rightarrow H$, duality invariance then requires:
\begin{equation}
\star H_{\mu\nu}=-\tensor{\chi}{_{\mu\nu}^{\alpha\beta}} F_{\alpha\beta}\, , \quad \star F_{\mu\nu}=\tensor{\chi}{_{\mu\nu}^{\alpha\beta}} H_{\alpha\beta}\, .
\end{equation}
Applying the star operator in both sides,
\begin{equation}
 H_{\mu\nu}=- \tensor{(\star \chi)}{_{\mu\nu}^{\alpha\beta}} F_{\alpha\beta}\, , \quad F_{\mu\nu}=\tensor{(\star \chi)}{_{\mu\nu}^{\alpha\beta}} H_{\alpha\beta}\, .
 \label{eq:dualinvlinear}
\end{equation}
Substituting the second equation into the first one,
\begin{equation}
H_{\mu\nu}=- \tensor{(\star \chi)}{_{\mu\nu}^{\alpha\beta}}\tensor{(\star \chi)}{_{\alpha \beta}^{\rho \sigma}} H_{\rho \sigma}\,,
\end{equation}
so \eqref{eq:chiinv} must hold. In order to prove sufficiency, we must ensure that the constitutive relation and the Einstein's equations remain invariant. The constitutive relation is easily seen to be invariant, since the equation \eqref{eq:constlinear0} together with its inverse can be written in a manifestly duality-invariant way as follows:
\begin{equation}\label{eq:constlinear}
	\tensor{\sigma}{^{A}_{B}}\mathcal{F}^B_{\mu\nu}=\tensor{(\star\chi)}{_{\mu\nu}^{\alpha\beta}} \mathcal{F}^A_{\alpha\beta}
\end{equation}
where $\mathcal{F}^A$ is the vector of 2-forms

\begin{equation}
	\mathcal{F}^A=\begin{pmatrix}
		F\\
		H
	\end{pmatrix}
\end{equation}
and $\tensor{\sigma}{^{A}_{B}}$ is the symplectic matrix
\begin{equation}
	\tensor{\sigma}{^{A}_{B}}=\begin{pmatrix}
		0 & 1\\
	-1    & 0
	\end{pmatrix}
\end{equation}
The $\mathrm{SO}(2)$ invariance of this equation follows from that of $\tensor{\sigma}{^{A}_{B}}$, and also note that the equation is consistent because both operators $\tensor{\sigma}{^{A}_{B}}$  and $\tensor{(\star\chi)}{_{\mu\nu}^{\alpha\beta}}$ satisfy that their square is minus the identity. 

Let us now show that the Einstein's equations are invariant as well. Since we are assuming that $\chi$ depends algebraically on the curvature, these can be written as 
\begin{equation}
G_{\mu\nu}=\tensor{P}{_{(\mu}^{\rho \sigma \gamma}} R_{\nu) \rho \sigma \gamma}+2 \nabla^\sigma \nabla^\rho P_{(\mu| \sigma | \nu)\rho} -\frac{1}{2}g_{\mu \nu} (\chi^{\alpha\beta\rho\sigma}F_{\alpha\beta}F_{\rho\sigma})+2F_{( \mu|\alpha}\tensor{\chi}{_{|\nu)}^{\alpha\rho\sigma}}F_{\rho\sigma}\, ,
\end{equation}
where
\begin{equation}
P^{\mu\nu\rho\sigma}=\frac{\partial \chi^{\alpha\beta\gamma\lambda}}{\partial R_{\mu\nu\rho\sigma}}	F_{\alpha\beta}F_{\gamma\lambda}\, .
\end{equation}
The two last terms can be arranged in a duality-invariant fashion as 

\begin{equation}
 -\frac{1}{2}g_{\mu \nu} (\chi^{\alpha\beta\rho\sigma}F_{\alpha\beta}F_{\rho\sigma})+2F_{( \mu|\alpha}\tensor{\chi}{_{|\nu)}^{\alpha\rho\sigma}}F_{\rho\sigma}=-2F_{\langle \mu|\alpha}\star \tensor{H}{_{|\nu\rangle}^{\alpha}}=-\sigma_{AB}\tensor{\mathcal{F}}{^A_{\langle \mu|\alpha}}\star \tensor{\mathcal{F}}{^{B}_{|\nu\rangle}^{\alpha}}\, ,
\end{equation}
where $\sigma_{AB}$ has the same matrix form as $\sigma^{A}{}_B$. On the other hand, the tensor $P^{\mu\nu\rho\sigma}$ is also invariant. To see this, let us first rewrite it as follows, 

\begin{equation}
P^{\mu\nu\rho\sigma}=-\frac{\partial \chi^{\alpha\beta\gamma\lambda}}{\partial R_{\mu\nu\rho\sigma}}\tensor{(\star \chi)}{_{\gamma\lambda}^{\tau\epsilon}}F_{\alpha\beta}H_{\tau\epsilon}
\end{equation}
where we are using the inverse of \eqref{eq:constlinear0}. 
Now, differentiating \req{eq:chiinv} with respect to the curvature, it follows that 

\begin{equation}
	\frac{\partial\tensor{(\star\chi)}{_{\alpha\beta}^{\gamma\lambda}}}{\partial R_{\mu\nu\rho\sigma}}	\tensor{(\star\chi)}{_{\gamma\lambda}^{\tau\epsilon}}+\tensor{(\star\chi)}{_{\alpha\beta}^{\gamma\lambda}}\frac{\partial\tensor{(\star\chi)}{_{\gamma\lambda}^{\tau\epsilon}}}{\partial R_{\mu\nu\rho\sigma}}=0\, ,
\end{equation}
which is equivalent to

\begin{equation}
	\frac{\partial\tensor{\chi}{^{\alpha\beta\gamma\lambda}}}{\partial R_{\mu\nu\rho\sigma}}	\tensor{(\star\chi)}{_{\gamma\lambda}^{\tau\epsilon}}+
\frac{\partial\tensor{\chi}{^{\tau\epsilon\gamma\lambda}}}{\partial R_{\mu\nu\rho\sigma}}\tensor{(\star\chi)}{_{\gamma\lambda}^{\alpha\beta}}=0\, .
\end{equation}
This allows us to write the tensor $P$ in a manifestly duality-invariant form
\begin{equation}
P^{\mu\nu\rho\sigma}=-\frac{1}{2}\frac{\partial \chi^{\alpha\beta\gamma\lambda}}{\partial R_{\mu\nu\rho\sigma}}\tensor{(\star \chi)}{_{\gamma\lambda}^{\tau\epsilon}}\sigma_{AB}\tensor{\mathcal{F}}{^A_{\alpha\beta}} \tensor{\mathcal{F}}{^{B}_{\tau\epsilon}}=\frac{1}{2}\frac{\partial \chi^{\alpha\beta\gamma\lambda}}{\partial R_{\mu\nu\rho\sigma}}\tensor{\mathcal{F}}{^A_{\alpha\beta}} \tensor{\mathcal{F}}{^{A}_{\gamma\lambda}}\, ,
\end{equation}
where in the last equality we used \req{eq:constlinear} and \req{eq:chiinv} in order to simplify the result. In sum, all the equations can be written as

\begin{align}
\notag
G^{\mu\nu}=-&\sigma_{AB}\tensor{\mathcal{F}}{^A^{\langle \mu|\alpha}}\star \tensor{\mathcal{F}}{^{B}^{|\nu\rangle}_{\alpha}}+ \frac{1}{2}\tensor{R}{^{(\mu}_{\rho \sigma \gamma}}\frac{\partial \chi^{\alpha\beta\lambda\tau}}{\partial R_{\nu)\rho\sigma\gamma}}\tensor{\mathcal{F}}{^A_{\alpha\beta}} \tensor{\mathcal{F}}{^{A}_{\lambda\tau}}\\
&+ \nabla_\sigma \nabla_\rho\left(\frac{\partial \chi^{\alpha\beta\lambda\tau}}{\partial R_{(\mu| \sigma | \nu)\rho}}\tensor{\mathcal{F}}{^A_{\alpha\beta}} \tensor{\mathcal{F}}{^{A}_{\lambda\tau}}\right) \, ,\\\notag
&\\
\tensor{\sigma}{^{A}_{B}}\mathcal{F}^B_{\mu\nu}=&\tensor{(\star\chi)}{_{\mu\nu}^{\alpha\beta}} \mathcal{F}^A_{\alpha\beta}\, ,\\\notag
&\\
d\mathcal{F}^A=&0\, .
\end{align}
Thus, we have proven that any tensor $\chi_{\mu\nu\rho\sigma}$ satisfying \req{eq:chiinv} defines a duality-invariant theory given by \req{eq:linearth}. To complete this section, let us characterize the tensors satisfying that property.  First, we note that, making use of the properties of the Levi-Civita tensor, the equation \req{eq:chiinv} can be rewritten as 

\begin{equation}
6\tensor{\chi}{_{[\alpha\beta}^{\alpha\beta}}\tensor{\chi}{_{\mu\nu]}^{\rho\sigma}}=\tensor{\delta}{_{[\mu}^{[\rho}}\tensor{\delta}{_{\nu]}^{\sigma]}}\, .
\end{equation}
This is a quadratic tensor equation that admits infinitely many solutions that cannot be written explicitly. However, since we are interested in theories that reduce to Einstein-Maxwell's at low energies, the tensor $\chi_{\mu\nu\rho\sigma}$ should reduce to the indentity when the curvature is small, and we can expand it as
\begin{equation}
\tensor{\chi}{_{\mu\nu}^{\rho\sigma}}=\tensor{\delta}{_{[\mu}^{[\rho}}\tensor{\delta}{_{\nu]}^{\sigma]}}+\sum_{n=1}^{\infty}\ell^{2n}\tensor{\chi}{^{(n)}_{\mu\nu}^{\rho\sigma}}\, ,
\end{equation}
where $\ell$ is a length scale and each $\tensor{\chi}{^{(n)}_{\mu\nu}^{\rho\sigma}}$ contains $2n$ derivatives of the metric. Inserting this expansion in the equation above yields the following relation for the $n$-th order tensor
\begin{equation}
6\tensor{\chi}{^{(n)}_{[\alpha\beta}^{\alpha\beta}}\tensor{\delta}{_{\mu}^{\rho}}\tensor{\delta}{_{\nu]}^{\sigma}}+6\tensor{\chi}{^{(n)}_{[\mu\nu}^{\rho\sigma}}\tensor{\delta}{_{\alpha}^{\alpha}}\tensor{\delta}{_{\beta]}^{\beta}}+6\sum_{p=1}^{n-1}\tensor{\chi}{^{(p)}_{[\alpha\beta}^{\alpha\beta}}\tensor{\chi}{^{(n-p)}_{\mu\nu]}^{\rho\sigma}}=0\, ,
\end{equation}
and after expanding the antisymmetrization in the first two terms, we can write this as follows, 

\begin{equation}\label{eq:eqchi}
2\tensor{\chi}{^{(n)}_{\mu\nu}^{\rho\sigma}}-4\tensor{\chi}{^{(n)}_{\alpha[\mu}^{\alpha[\rho}}\tensor{\delta}{_{\nu]}^{\sigma]}}+\tensor{\chi}{^{(n)}_{\alpha\beta}^{\alpha\beta}}\tensor{\delta}{_{[\mu}^{\rho}}\tensor{\delta}{_{\nu]}^{\sigma}}=-6\sum_{p=1}^{n-1}\tensor{\chi}{^{(p)}_{[\alpha\beta}^{\alpha\beta}}\tensor{\chi}{^{(n-p)}_{\mu\nu]}^{\rho\sigma}}\, .
\end{equation}
Now, this is an inhomogeneous linear tensor equation for $\tensor{\chi}{^{(n)}_{\mu\nu}^{\rho\sigma}}$, and so, the general solution can be expressed as the sum of a particular solution plus the general solution of the associated homogeneous equation. The latter reads

\begin{equation}
2\tensor{\chi}{_{\rm h}^{(n)}_{\mu\nu}^{\rho\sigma}}-4\tensor{\chi}{_{\rm h}^{(n)}_{\alpha[\mu}^{\alpha[\rho}}\tensor{\delta}{_{\nu]}^{\sigma]}}+\tensor{\chi}{_{\rm h}^{(n)}_{\alpha\beta}^{\alpha\beta}}\tensor{\delta}{_{[\mu}^{\rho}}\tensor{\delta}{_{\nu]}^{\sigma}}=0\, ,
\end{equation}
and taking the trace in $\nu\sigma$ we have 

\begin{equation}
\frac{1}{2}\tensor{\chi}{_{\rm h}^{(n)}_{\alpha\beta}^{\alpha\beta}}\tensor{\delta}{_{\mu}^{\rho}}=0\, .
\end{equation}
Therefore, we have $\tensor{\chi}{_{\rm h}^{(n)}_{\alpha\beta}^{\alpha\beta}}=0$ and the general solution of the homogeneous equation can be expressed as
\begin{equation}
\tensor{\chi}{_{\rm h}^{(n)}_{\mu\nu}^{\rho\sigma}}=\tensor{\mathcal{T}}{^{(n)}_{[\mu}^{[\rho}}\tensor{\delta}{_{\nu]}^{\sigma]}}\, ,\quad \text{where} \quad \tensor{\mathcal{T}}{^{(n)}_{\mu}^{\mu}}=0\, .
\end{equation}
This is, the solution is characterized by an arbitrary traceless (and symmetric) tensor $\mathcal{T}^{(n)}_{\mu\nu}$. Now, coming back to the inhomogeneous equation, we realize that, since the trace of the left-hand-side of \req{eq:eqchi} is proportional to the identity, a necessary condition in order for a solution to exist is that 
\begin{equation}\label{eq:consistency}
\sum_{p=1}^{n-1}\tensor{\chi}{^{(p)}_{[\alpha\beta}^{\alpha\beta}}\tensor{\chi}{^{(n-p)}_{\mu\nu]}^{\rho\nu}}\propto \tensor{\delta}{_{\mu}^{\rho}}\, .
\end{equation}
However, this is guaranteed by the following property satisfied by all tensors $Q^{(1)}{}_{\mu \nu \rho \sigma}$ and $Q^{(2)}{}_{\mu \nu \rho \sigma}$ which are antisymmetric in the indices $\{\mu \nu\}$ and $\{\rho \sigma\}$:
\begin{equation}
Q^{(1)}{}_{[\alpha \beta}{}^{\alpha \beta} Q^{(2)}{}_{\mu \lambda]\nu}{}^\lambda +Q^{(2)}{}_{[\alpha \beta}{}^{\alpha \beta} Q^{(1)}{}_{\mu \lambda]\nu}{}^\lambda =\frac{1}{2}Q^{(1)}{}_{[\alpha \beta}{}^{\alpha \beta} Q^{(2)}{}_{\gamma \lambda]}{}^{\gamma\lambda} g_{\mu \nu}\,.
\label{eq:propchuli}
\end{equation}
This property is proven by expanding the antisymmetrization in the identity $Q^{(1)}{}_{[\alpha \beta}{}^{\alpha \beta} Q^{(2)}{}_{\gamma \lambda}{}^{\gamma\lambda} g_{\mu] \nu}=0$. Since the summation in \eqref{eq:consistency} is symmetric in the exchange of $\chi^{(p)}$ and $\chi^{(n-p)}$ and every such $\chi^{(p)}$ and $\chi^{(n-p)}$ have the required symmetries for \eqref{eq:propchuli} to hold, we conclude that \eqref{eq:consistency} is always satisfied.

Hence, taking into account \eqref{eq:propchuli}, one can see that the general solution to \eqref{eq:eqchi} reads,

\begin{equation}
\tensor{\chi}{^{(n)}_{\mu\nu}^{\rho\sigma}}=\tensor{\mathcal{T}}{^{(n)}_{[\mu}^{[\rho}}\tensor{\delta}{_{\nu]}^{\sigma]}}-3\sum_{p=1}^{n-1}\tensor{\chi}{^{(p)}_{[\alpha\beta}^{\alpha\beta}}\tensor{\chi}{^{(n-p)}_{\mu\nu]}^{\rho\sigma}}\, ,
\end{equation}
and thus it is determined by a set of traceless symmetric tensors $\{\mathcal{T}^{(n)}_{\mu\nu}\}_{n\ge 1}$.
Once these tensors are specified, one can compute the tensor $\tensor{\chi}{_{\mu\nu}^{\rho\sigma}}$ at arbitrary orders by using this recursive relation. Note that even when the set of non-vanishing $\mathcal{T}^{(n)}_{\mu\nu}$ tensors is finite, the series contains always an infinite number of terms. Even though there seems to be no simple way of writing the general term in this expansion, at least we have a systematic way of computing terms of arbitrarily high order --- a task that becomes much more involved in theories that are non-linear in $F$.

\section{Field Redefinitions}\label{sec:redef}
In section~\ref{sec:dualityrot}, we obtained the most general duality-invariant action up to the eight derivative level. This can be understood as the truncation of an exactly duality-invariant theory that necessarily contains an infinite tower of higher-derivative terms. But at the same time, it can be interpreted as the effective field theory of some underlying UV-complete theory that respects electromagnetic duality. From this point of view, it is very natural to consider field redefinitions of the metric and vector fields, since these correspond simply to different choices of renormalization schemes of the hypothetical quantum theory, leaving the physics invariant.  
One is of course free to redefine the variables of any given theory, and if the original theory possess a symmetry, so must the transformed one. However, if the change of variables is not invariant under that symmetry, then the new action will not be manifestly symmetric. 
The goal of this section is to investigate this question in the case of the duality-invariant theories under consideration. We would like to find field redefinitions that map these theories into other of the same class, with the final aim of removing as many parameters as we can from the Lagrangian.

Let us start by writing down our duality-invariant action up to the six-derivative level
\begin{equation}
S=\frac{1}{16\pi G}\int d^4x\sqrt{|g|}\left\{R-F^2+\ell^2\mathcal{L}_{(4)}+\ell^4\left(\mathcal{L}_{(6)}^{\rm H}-\frac{1}{8}\frac{\partial \mathcal{L}_{(4)}}{\partial \tensor{F}{^{\alpha\beta}}}\frac{\partial \mathcal{L}_{(4)}}{\partial \tensor{F}{_{\alpha\beta}}}\right)+\mathcal{O}(\ell^6)\right\}\, ,
\end{equation}
where we recall that $\mathcal{L}_{(4)}$ reads
\begin{equation}
\mathcal{L}_{(4)}=\alpha_1 T_{\mu\nu}T^{\mu\nu}+\alpha_2 T_{\mu\nu}R^{\mu\nu}+ \alpha_3 \mathcal{X}_4+\alpha_4 R_{\mu \nu} R^{\mu \nu}+\alpha_5 R^2\, ,
\end{equation}
and $\mathcal{L}_{(6)}^{\rm H}$ is given by \req{eq:homl6sol}. Let us then consider a redefinition of the metric of the form

\begin{equation}\label{eq:redef1}
g_{\mu \nu}\rightarrow g_{\mu \nu}+\ell^{2} h_{\mu \nu}\, ,
\end{equation}
where $h_{\mu\nu}$ is some symmetric 2-derivative tensor.  Performing such field redefinition, expanding in powers of $\ell^2$, and neglecting total derivatives, the action $S$ undergoes the transformation

\begin{align}\notag
S'&=S+\frac{1}{16 \pi G}\int d^4 x  \sqrt{\vert g \vert}\Bigg\{-\ell^2 h^{\mu\nu}(G_{\mu\nu}-2T_{\mu\nu})+\ell^4\bigg[\frac{1}{8}\left(2h_{\alpha\beta}h^{\alpha\beta}-h^2\right)(R-F^2)\\\notag
& \left(h^{\mu\alpha}\tensor{h}{^{\nu}_{\alpha}}-\frac{1}{2}h h^{\mu\nu}\right)(G_{\mu\nu}-2T_{\mu\nu}) -h^{\mu\nu}h^{\alpha\beta}F_{\mu\alpha}F_{\nu\beta}-\frac{1}{4}\nabla_{\mu}h_{\alpha\beta}\nabla^{\mu}h^{\alpha\beta}+\frac{1}{4}\nabla_{\mu}h\nabla^{\mu}h\\\notag
&+\frac{1}{2}\nabla_{\mu}h^{\alpha\beta}\nabla_{\beta}\tensor{h}{_{\alpha}^{\mu}}-\frac{1}{2}\nabla_{\mu}h^{\mu\alpha}\nabla_{\alpha}h-h^{\mu\nu}\bigg(\tensor{R}{^{\lambda}_{\mu}}\frac{\partial \mathcal{L}_{(4)}}{\partial R^{\nu\lambda}}+\frac{1}{2}g_{\mu\nu}\nabla^{\alpha}\nabla^{\beta}\frac{\partial \mathcal{L}_{(4)}}{\partial R^{\alpha\beta}}\\
&-\nabla^{\alpha}\nabla_{\mu}\frac{\partial \mathcal{L}_{(4)}}{\partial R^{\nu\alpha}}+\frac{1}{2}\nabla^2\frac{\partial \mathcal{L}_{(4)}}{\partial R^{\mu\nu}}-\frac{1}{2}g_{\mu\nu}\mathcal{L}_{(4)}+F_{ \mu\alpha}\frac{\partial \mathcal{L}_{(4)}}{\partial \tensor{F}{^{\nu}_{\alpha}}}\bigg)\bigg]\Bigg\}\,.
\end{align}

\noindent
Then, the idea is to choose a tensor $h_{\mu\nu}$ that simplifies the Lagrangian. Note that, in the four-derivative Lagrangian, the redefinition introduces terms proportional to $T_{\mu\nu}$, and this allows one to remove all the terms depending on the Maxwell stress tensor.  
This is achieved by the redefinition
\begin{equation}
h_{\mu \nu}=-\frac{\alpha_1}{2} T_{\mu \nu}-\frac{\alpha_1+2\alpha_2}{4} \hat{R}_{\mu \nu}+\sigma R g_{\mu \nu} \,,
\end{equation}
where $\hat{R}_{\mu \nu}=R_{\mu \nu}-\frac{1}{4}g_{\mu \nu} R$ and $\sigma$ is a free parameter. This has the following effect on $\mathcal{L}_{(4)}$:

\begin{equation}
\mathcal{L}'_{(4)}=\alpha_3 \mathcal{X}_4+\alpha_4' R_{\mu \nu} R^{\mu \nu}+\alpha_5' R^2\, ,
\end{equation}
where
\begin{equation}\label{eq:alpha45}
\alpha_4'=\alpha_4+\frac{\alpha_1}{4}+\frac{\alpha_2}{2}\, ,\qquad \alpha_5'=\alpha_5-\frac{\alpha_1}{16}-\frac{\alpha_2}{8}+\sigma\, .
\end{equation}
In this way, we have removed all the terms containing field strengths in the four-derivative Lagrangian, and hence, the new theory is obviously duality-invariant at that order. However, this redefinition also affects the 6- and higher-derivative Lagrangians, so we must investigate if duality is also preserved at that higher orders. 
Let us note that, in this new frame, the Maxwell stress tensor $T_{\mu\nu}$ is invariant to order $\ell^2$ under a duality rotation due to the absence of field strengths in $\mathcal{L}_{(4)}$. Thus, it follows that the redefinition \req{eq:redef1} is invariant to order $\ell^4$, and hence once would expect the transformed theory to be also duality-invariant at that order. 
We recall that the presence of terms with $T_{\mu\nu}$ in $\mathcal{L}_{(4)}$ induces inhomogenous terms in $\mathcal{L}_{(6)}$ that are required by duality. Now, since the redefinition removes the terms with $T_{\mu\nu}$ in $\mathcal{L}_{(4)}$, our intuition is that it should also remove the inhomogeneous terms associated with these, since otherwise duality symmetry would be broken. In fact, this is almost exactly what happens. 

After a somewhat lengthy computation, we arrive at the following expression for the transformed action up to $\mathcal{O}(\ell^4)$:

\begin{align}\notag
S'=&\frac{1}{16\pi G}\int d^4x\sqrt{|g|}\Bigg\{R-F^2+\ell^2\left(\alpha_3 \mathcal{X}_4+\alpha_4' R_{\mu \nu} R^{\mu \nu}+\alpha_5' R^2\right)+\\
&\ell^4\left( \left(\mathcal{L}_{(6)}^{\rm H}\right)'+\mathcal{O}\left(\nabla T\right)-\frac{\alpha_1^2}{64}\mathcal{E}_{\mu\nu}\mathcal{E}^{\langle\mu\nu\rangle} F^2-\frac{ \alpha_1^2}{16}\mathcal{E}^{\langle\mu\nu\rangle} \mathcal{E}^{\langle\alpha\beta\rangle}  F_{\alpha \mu} F_{\beta \nu}\right)+\mathcal{O}(\ell^6)\Bigg\}\, ,
\end{align}
where $\mathcal{E}_{\mu\nu}=G_{\mu\nu}-2T_{\mu\nu}$ are the zeroth order equations of motion. 
Here, $ \left(\mathcal{L}_{(6)}^{\rm H}\right)'$ is the homogeneous duality-invariant Lagrangian \req{eq:homl6sol} with renormalized couplings, while $\mathcal{O}\left(\nabla T\right)$ represents new terms that, besides depending on the curvature and the Maxwell stress tensor, also contain derivatives of the latter. These terms were not included in the original action because, as we argued, they cannot arise in the truncation of a theory that is \emph{exactly} invariant under duality rotations. 
However, the field redefinitions we are considering are defined only perturbatively and they generically introduce this type of terms, which indeed preserve duality in a perturbative sense. In any case, as we show below, one can easily get rid of them. 
Thus, the only problematic terms are those that depend explicitly on the field strength. Apparently, these terms break duality invariance, but a closer look reveals that this is not so. Indeed, since they are proportional to the square of the zeroth-order equations of motion, this implies that, on-shell, their contribution is of order $\ell^6$, and hence the equations of the new theory are actually invariant under duality rotations at order $\ell^4$. Moreover, we can simply remove those terms from the action by performing an additional redefinition of the metric, 
\begin{equation}\label{eq:redef2}
g_{\mu \nu}\rightarrow g_{\mu \nu}+\ell^{4} h^{(4)}_{\mu \nu}\, .
\end{equation}
This yields,

\begin{align}\notag
S''=&\frac{1}{16\pi G}\int d^4x\sqrt{|g|}\Bigg\{R-F^2+\ell^2\left(\alpha_3 \mathcal{X}_4+\alpha_4' R_{\mu \nu} R^{\mu \nu}+\alpha_5' R^2\right)+\\
&\ell^4\left( \left(\mathcal{L}_{(6)}^{\rm H}\right)'+\mathcal{O}\left(\nabla T\right)-\frac{\alpha_1^2}{64}\mathcal{E}_{\mu\nu}\mathcal{E}^{\langle\mu\nu\rangle} F^2-\frac{ \alpha_1^2}{16}\mathcal{E}^{\langle\mu\nu\rangle} \mathcal{E}^{\langle\alpha\beta\rangle}  F_{\alpha \mu} F_{\beta \nu}-\mathcal{E}^{\mu\nu}h^{(4)}_{\mu \nu}\right)+\mathcal{O}(\ell^6)\Bigg\}\, ,
\end{align}
so that, by choosing,

\begin{equation}
h^{(4)}_{\mu \nu}=-\frac{\alpha_1^2}{64}\mathcal{E}_{\langle\mu\nu\rangle} F^2-\frac{ \alpha_1^2}{16}\mathcal{E}^{\langle\alpha\beta\rangle}  F_{\alpha \langle\mu|} F_{\beta |\nu\rangle}\, ,
\end{equation}
we cancel those terms. Note that, again, this redefinition is of order $\ell^6$ on-shell, and hence it is trivially invariant at order $\ell^4$. In fact, it makes sense that duality invariance is preserved only on-shell, since this is a symmetry of the equations of motion, not of the action. 

We can now perform additional $\mathcal{O}(\ell^4)$ redefinitions in order to simplify the six-derivative Lagrangian. Since these introduce terms of the form  $G^{\mu\nu}h^{(4)}_{\mu \nu}-2T^{\mu\nu}h^{(4)}_{\mu \nu}$, this means that we can remove all the terms depending on the stress-energy tensor $T_{\mu\nu}$, or more precisely, we can simply perform the replacement $T_{\mu\nu}\rightarrow G_{\mu\nu}/2$. Note that this works, too, for the terms that contain derivatives of $T_{\mu\nu}$, since, by integration by parts, it is always possible to remove the derivatives from one stress-energy tensor, and then the replacement above can be applied. 
Since all the pure-metric higher-derivative terms were already included in the original action, these redefinitions have the only effect of renormalizing their couplings while removing the dependence on $T_{\mu\nu}$ --- and hence, $F_{\mu\nu}$ --- in the six-derivative Lagrangian.

Thus, we have reached a quite remarkable result: to the six-derivative level, the most general duality-invariant extension of Einstein-Maxwell theory is equivalent to the most general higher-order gravity minimally coupled to Maxwell theory. 
Explicitly, the action reads\footnote{Note that the numbering in the $\beta_i$ couplings is different from the one used in section~\ref{sec:all8d}.}

\begin{align}\notag
S'''=&\frac{1}{16\pi G}\int d^4x\sqrt{|g|}\Bigg\{R-F^2+\ell^2\left(\alpha_3 \mathcal{X}_4+\alpha_4' R_{\mu \nu} R^{\mu \nu}+\alpha_5' R^2\right)+\\\notag
&\ell^4\left( \beta_1\tensor{R}{_{\mu}^{\rho}_{\nu}^{\sigma}}\tensor{R}{_{\rho}^{\alpha}_{\sigma}^{\beta}}\tensor{R}{_{\alpha}^{\mu}_{\beta}^{\nu}}+\beta_2 \tensor{R}{_{\mu \nu}^{\rho \sigma}}\tensor{R}{_{\rho \sigma}^{\alpha \beta}}\tensor{R}{_{\alpha \beta}^{\mu\nu}}+\beta_3 \tensor{R}{_{\mu \nu \rho \sigma}}\tensor{R}{^{\mu \nu \rho}_{\alpha}}R^{\sigma \alpha}\right.\\ \notag
&\left.+\beta_4\tensor{R}{_{\mu \nu \rho \sigma}}\tensor{R}{^{\mu \nu \rho \sigma}}R+\beta_5\tensor{R}{_{\mu \nu \rho \sigma}}\tensor{R}{^{\mu \rho}}\tensor{R}{^{\nu \sigma}}+\beta_6\tensor{R}{_{\mu}^{\nu}}\tensor{R}{_{\nu}^{\rho}}\tensor{R}{_{\rho}^{\mu}}+\beta_7R_{\mu \nu }R^{\mu \nu }R\right.\\
&\left.+\beta_8R^3+\beta_9 \nabla_{\sigma}R_{\mu \nu} \nabla^{\sigma}R^{\mu\nu}+\beta_{10}\nabla_{\mu}R\nabla^{\mu}R\right)+\mathcal{O}(\ell^6)\Bigg\}\, ,
\label{eq:Sppp}
\end{align}
where we are including all the six-derivative Riemann invariants modulo total derivatives \cite{Fulling:1992vm}.  However, we can still decrease the number of terms in this action. To begin with, not all the cubic invariants are independent, since they satisfy two constraints \cite{Fulling:1992vm}, and this allows us to set, for instance $\beta_1=\beta_3=0$. On the other hand, there are residual redefinitions of the metric that cancel some of these curvature terms without introducing field strengths.  In the four-derivative Lagrangian, we recall that $\alpha_5'$ is given by \req{eq:alpha45}, where $\sigma$ is arbitrary. Thus, we are free to choose $\sigma=-\alpha_5+\frac{\alpha_1}{16}+\frac{\alpha_2}{8}$, so that $\alpha_5'=0$. In the case of the six-derivative terms, notice that a redefinition of the form $g_{\mu \nu}\rightarrow g_{\mu \nu}(1+\ell^{4} h^{(4)})$ adds the term $\ell^{4}h^{(4)}R$ to the Lagrangian. Hence, we can cancel all the terms that contain at least one Ricci scalar, and thus we can set $\beta_4=\beta_7=\beta_8=\beta_{10}=0$. Finally, since we can transform all Ricci tensors into Maxwell stress-energy tensors via metric redefinitions (up to terms that involve Ricci scalars, and that hence can be removed), we have the map 
\begin{equation}
\tensor{R}{_{\mu}^{\nu}}\tensor{R}{_{\nu}^{\rho}}\tensor{R}{_{\rho}^{\mu}}\rightarrow 8 \tensor{T}{_{\mu}^{\nu}}\tensor{T}{_{\nu}^{\rho}}\tensor{T}{_{\rho}^{\mu}}=0\, .
\end{equation}
Therefore, we can also set $\beta_6=0$. In sum, the action is simplified to\footnote{With respect to \req{eq:Sppp}, we are relabeling the couplings as $\alpha_3 \rightarrow \alpha_1$, $\alpha_4' \rightarrow \alpha_2$, $\beta_2 \rightarrow \beta_1$, $\beta_5 \rightarrow \beta_2$ and $\beta_9 \rightarrow \beta_3$.}

\begin{align}\notag
S=&\frac{1}{16\pi G}\int d^4x\sqrt{|g|}\Bigg\{R-F^2+\ell^2\left(\alpha_1 \mathcal{X}_4+\alpha_2 R_{\mu \nu} R^{\mu \nu}\right)+\\
&\ell^4\left(\beta_1 \tensor{R}{_{\mu \nu}^{\rho \sigma}}\tensor{R}{_{\rho \sigma}^{\alpha \beta}}\tensor{R}{_{\alpha \beta}^{\mu\nu}}+\beta_2\tensor{R}{_{\mu \nu \rho \sigma}}\tensor{R}{^{\mu \rho}}\tensor{R}{^{\nu \sigma}}+\beta_3 \nabla_{\sigma}R_{\mu \nu} \nabla^{\sigma}R^{\mu\nu}\right)+\mathcal{O}(\ell^6)\Bigg\}\, ,
\label{eq:EFT6}
\end{align}
and it only contains five independent operators, of which one is topological.  Thus, duality invariance together with field redefinitions is a powerful tool that removes most of the higher-order terms one can include in the action.

One may wonder if this result extends to even higher-derivative terms. As we have discussed, at order $n$, the Lagrangian of a duality-invariant theory can be decomposed as $\mathcal{L}_{(2n)}=\mathcal{L}_{(2n)}^{\rm H}+\mathcal{L}_{(2n)}^{\rm IH}$, where $\mathcal{L}_{(2n)}^{\rm H}$ contains new independent terms that depend on $F_{\mu\nu}$ only through $T_{\mu\nu}$, and $\mathcal{L}_{(2n)}^{\rm IH}$ is determined by the lower-order Lagrangians. Now, if up to order $n-1$ we have been able to cancel all the higher-order terms containing $F_{\mu\nu}$ via field redefinitions, we expect that those redefinitions also cancel  $\mathcal{L}_{(2n)}^{\rm IH}$
up to terms which are duality-invariant at that order, and that hence depend on $T_{\mu\nu}$. This is precisely what happened with $\mathcal{L}_{(6)}^{\rm IH}$ when we performed the redefinition that cancels the $F$-dependent terms in $\mathcal{L}_{(4)}$. If this is the case, then the corresponding transformed Lagrangian $\mathcal{L}_{(2n)}'$ will depend on $F_{\mu\nu}$ only through $T_{\mu\nu}$ and therefore there is an additional metric redefinition of $2n-2$ derivatives that maps that Lagrangian into a pure gravity one. By induction, one would conclude that this process can be carried out to all orders. As we have seen, things are not so simple, since in this process of field redefinitions duality is only preserved on-shell, which adds some additional complications to the argument. Still, the evidence gathered so far makes us confident to propose the following
\begin{conjecture}
Any duality-invariant theory that allows for a derivative expansion around Einstein-Maxwell theory is perturbatively equivalent to Maxwell theory minimally coupled to a higher-derivative gravity at any order in the derivative expansion.  
\end{conjecture}
As a further support of this conjecture it would interesting to carry out the explicit computation for the eight-derivative terms, although this is a highly challenging task that entails the computation of the third variation of the Einstein-Maxwell action as well as the second variation of the Lagrangian $\mathcal{L}_{(4)}$. Interestingly, if the conjecture holds, then it means that, when coupled to Einstein gravity, non-linear duality invariant electromagnetic theories such as Born-Infeld theory are in fact perturbatively equivalent to Maxwell theory coupled to higher-derivative gravity. Another interesting question is whether one can find a fully non-perturbative equivalence. 

Let us close this section with a few additional comments. In our approach to field redefinitions we have not included a cosmological constant since it introduces a new length scale besides $\ell$. This makes the computations more involved since now a redefinition of order $n$ affects linearly the Lagrangians of order $n$ and $n+1$ and new scales $\Lambda \ell^{2n}$ are generated. However, upon the assumption that $\Lambda\ell^2<<1$ we believe our qualitative results for the asymptotically flat case can be applied as well --- see \cite{Bueno:2019ltp} for an argument in a similar situation. 
Finally, we have been able to map any duality-invariant action to a theory that only contains metric higher-derivative terms, so one may wonder if one can make an analogous transformation to a frame in which the theory takes the form of Einstein gravity coupled to non-linear electrodynamics. However, this is not the case since terms that depend explicitly on the Riemann curvature such as $\tensor{R}{_{\mu \nu}^{\rho \sigma}}\tensor{R}{_{\rho \sigma}^{\alpha \beta}}\tensor{R}{_{\alpha \beta}^{\mu\nu}}$ or $R^{\mu\nu\alpha\beta}T_{\mu\alpha}T_{\nu\beta}$ cannot be mapped into terms containing only field strengths. Therefore, the action \req{eq:EFT6} represents possibly the simplest form for a general six-derivative duality-invariant theory.

\section{Black holes}\label{sec:BHs}
As a final application of our results, in this section we study the spherically symmetric black hole solutions of \req{eq:EFT6}, which, as we have shown, is equivalent to the most general duality-invariant theory to the six-derivative level. 
A general static and spherically symmetric ansatz can be written as
\begin{align}
ds^2&=-N(r)^2f(r)dt^2+\frac{dr^2}{f(r)}+r^2\left(d\theta^2+\sin^2\theta d\phi^2\right)\, ,\\
A&= A_t(r) dt-p \cos\theta d\phi\, ,
\end{align}
where $N(r)$, $f(r)$ and $A_t(r)$ are functions and $p$ is a constant. The field strength $F$ then reads

\begin{equation}
F=-A_t'(r)dt\wedge dr+p \sin\theta d\theta\wedge d\phi\, ,
\label{eq:fsbhsol}
\end{equation}
and the Maxwell's equations, which do not receive any corrections, read simply $d\star F=0$. The magnetic part of $F$ automatically satisfies this equation, while $A_t$ satisfies
\begin{equation}
\frac{d}{dr}\left(\frac{r^2A_t'}{N}\right)=0\, ,\quad \Rightarrow \quad A_t'=-\frac{N q}{r^2}\, ,
\end{equation}
where $q$ is an integration constant. This fully characterizes the field strength in terms of the parameters $q$ and $p$, which are the electric and magnetic charges as defined by

\begin{equation}
q=\frac{1}{4\pi}\int_{\Sigma} \star F\, ,\qquad p=\frac{1}{4\pi}\int_{\Sigma} F\ , 
\end{equation}
where $\Sigma$ is any surface that encloses $r=0$.  
Finally, Einstein's equations can be easily solved if one assumes a perturbative expansion of the functions $N(r)$ and $f(r)$ as 
\begin{equation}
N(r)=\sum_{n=0}^{\infty}N_n(r) \ell^{2n}\, ,\qquad f(r)=\sum_{n=0}^{\infty}f_n(r) \ell^{2n}\, .
\end{equation}
In this way, at each order the functions $\{f_n,N_n\}$ with $n>0$ satisfy the inhomogeneous linearized Einstein's equations, whose resolution is straightforward. The integration constants in this process are fixed so that $N(r\rightarrow \infty)=1$ and $f(r\rightarrow \infty)= 1-2M/r+\ldots$. Then, the result reads

\begin{align}
\notag
N(r)^2=&1+\frac{2 \ell^2 Q^2 \alpha _2}{r^4}+\ell^4 \left(Q^2\alpha _2^2\left(-\frac{80 Q^2}{r^8}+\frac{128 M
   }{r^7}-\frac{48}{r^6}\right)+Q^2\beta_2\left(-\frac{60 Q^2}{r^8}+\frac{32 M}{r^7}\right)  \right.\\\label{eq:npertur}
&\left.+ \beta_1\left(-\frac{612
   Q^4}{r^8}+\frac{4992 M Q^2}{7 r^7}-\frac{216 M^2}{r^6}\right)+Q^2\beta _3\left(-\frac{150
   Q^2}{r^8}+\frac{192 M }{r^7}-\frac{48 }{r^6}\right) \right)\, , \\
\notag
f(r)=& 1-\frac{2 M}{r}+\frac{Q^2}{r^2}+\ell^2Q^2\alpha _2 \left(-\frac{12 Q^2}{5 r^6}+\frac{6 M}{r^5}-\frac{4}{r^4}\right)\\\notag
& +\ell^4 \Bigg[ \alpha_2^2Q^2\left(\frac{1408 Q^4}{15
   r^{10}}-\frac{351 M Q^2}{r^9}+\frac{320 M^2 +\frac{1192 Q^2}{7}}{r^8}-\frac{304 M
   }{r^7}+\frac{72}{r^6}\right) \\\notag
&+\beta_1\left(\frac{1724 Q^6}{3
   r^{10}}-\frac{1884 M Q^4}{r^9}+\frac{\frac{11064 M^2 Q^2}{7}+672 Q^4}{r^8}-\frac{8
   \left(49 M^3+96 M Q^2\right)}{r^7}+\frac{216 M^2}{r^6}\right) \\\notag
&+\beta_2Q^2\left(\frac{521 Q^4}{9 r^{10}}-\frac{158 M Q^2}{r^9}+\frac{72 M^2+68
   Q^2}{r^8}-\frac{40 M}{r^7}\right) \\
&+ \beta_3Q^2\left(\frac{1472 Q^4}{9
   r^{10}}-\frac{566 M Q^2}{r^9}+\frac{464 M^2 +\frac{1752 Q^2}{7}}{r^8}-\frac{384 M
   }{r^7}+\frac{72}{r^6}\right)\Bigg]\, ,
\end{align}
where 
\begin{equation}
Q=\sqrt{q^2+p^2}\, .
\end{equation}
This solution obviously reduces to the (dyonic) Reissner-Nordstr\"om solution when all the couplings are set to zero. Note also that in the case of vanishing charge the only non-trivial correction is the one associated to $\beta_1$, since the rest of the interactions involve Ricci curvature. Let us now study some properties of this solution. 
When the charge-to-mass ratio is small enough, it represents a black hole, whose horizon $r_+$ corresponds to the largest root of the equation $f(r_+)=0$. Since near extremality the zeroth-order solution has a double root, one has to be careful when studying the solutions to the corrected equation. Thus, let us first consider the case in which we are far from extremality, meaning that $0<M^2-Q^2>>\ell^2$. Note that, if $Q>>\ell$, this condition still allows us to get relatively close to extremality, in the sense that we can have $M^2-Q^2<<Q^2$, but not \emph{too} close. In this regime the horizon radius receives corrections of order $\ell^2$, and it is not difficult to solve the equation $f(r_+)=0$ perturbatively in $\ell$ in order to get

\begin{align}
\nonumber
r_+=&M (1+\zeta )+\frac{\ell^2 \left(1+3 \zeta -4 \zeta ^2\right) \alpha _2}{5 M \zeta 
   (1+\zeta )^2}-\frac{\ell^4 (1-\zeta )^2 \left(21+147 \zeta +773 \zeta ^2+1984 \zeta
   ^3\right) \alpha _2^2}{1050 M^3 \zeta ^3 (1+\zeta )^5}\\\nonumber
&+\frac{2 \ell^4 \left(4-39 \zeta
   +336 \zeta ^2-511 \zeta ^3\right) \beta _1}{21 M^3 \zeta  (1+\zeta )^5}+\frac{\ell^4 (1-\zeta ) (-1+7 \zeta ) (-1+13 \zeta ) \beta _2}{18 M^3 \zeta  (1+\zeta)^5}\\
&+\frac{\ell^4 (1-\zeta ) (5+\zeta  (35+464 \zeta )) \beta _3}{63 M^3 \zeta 
   (1+\zeta )^5}+\mathcal{O}(\ell^6)\, ,
\end{align}
where $\zeta$ is the ``extremality parameter''

\begin{equation}
\zeta=\sqrt{1-\frac{Q^2}{M^2}}\, ,
\end{equation}
which ranges from $1$ in the uncharged case to (near) $0$ at extremality. Note that the corrections seem to diverge when $\zeta\rightarrow 0$, but this is only because the assumption that the corrections are of order $\ell^2$ is no longer correct. Indeed, when $M^2-Q^2\sim \ell^2$ one can see that the leading correction to $r_+$ is of order $\ell$ rather than $\ell^2$. Thus, the expression above is reliable for $\zeta\gtrsim \ell/M$, which can in fact be very small. 

Let us then study what happens exactly at extremality. This is achieved when $r_+$ is a double root of $f$, and hence we also have the condition $f'(r_+)=0$. We find that the radius and mass at extremality read
\begin{align}\label{eq:rext}
r_+^{\rm ext}=&Q-\frac{3\ell^4 (4 \beta_1+\beta_2)}{2Q^3}+\mathcal{O}(\ell^6)\, ,\\
M^{\rm ext}=&Q-\frac{\alpha_2\ell^2}{5 Q}-\frac{\ell^4(3 \alpha_2^2+48\beta_1+7\beta_2+10 \beta_3)}{126 Q^3}+\mathcal{O}(\ell^6)\, .
\label{eq:Mext}
\end{align}

\subsection{Black hole thermodynamics}

Once we have found perturbatively the most general static and spherically solution to the theory defined by \eqref{eq:EFT6}, our next objective is to study the thermodynamics of the corresponding black hole solutions. To this aim, we are interested in several physical magnitudes, namely: the black hole mass $M$, its temperature $T$, its entropy $S$, and the electric and ``magnetic'' potentials at the horizon, $\Phi(r_+)$ and $\Psi(r_+)$ respectively\footnote{It is also possible to compute the on-shell action, which can be seen to be invariant under rotations of the electric and magnetic charges when appropriate boundary terms are included \cite{Deser:1996xu}.}. For convenience, we shall express all these quantities in terms of $r_+$ and $Q$. 

We begin by obtaining the black hole mass as a function $M=M(r_+,Q)$. This can be done by imposing the condition $f(r_+)=0$, and one gets
\begin{equation}
\begin{split}
M&=\frac{r_+}{2}+\frac{Q^2}{2r_+}+ \frac{Q^2 (3 Q^2-5 r_+^2) \alpha_2 \ell^2}{10 r_+^5}+\frac{\alpha_2^2 Q^4 \ell^4}{84 r_+^9} (7Q^2-9r_+^2)\\&+\frac{\beta_1  \ell^4 }{42 r_+^9}(-445 Q^6+909 Q^4 r_+^2-585 Q^2 r_+^4+105 r_+^6)- \frac{Q^2 \beta_2 \ell^4}{18 r_+^9}(28 Q^4-45 Q^2 r_+^2+18 r_+^4)\\&- \frac{Q^2 \beta_3 \ell^4}{126 r_+^9}(217 Q^4-459 Q^2 r_+^2+252 r_+^4)+\mathcal{O}(\ell^6)\,.
\label{eq:massrh}
\end{split}
\end{equation}
The black hole temperature is given by the general formula
\begin{equation}
T=\frac{f'(r_+)N(r_+)}{4\pi}\, ,
\end{equation}
and when expressed in terms of $r_+$ and $Q$ yields the following result,
\begin{equation}
\begin{split}
T&=\frac{r_+^2-Q^2}{4 \pi r_+^3}+\frac{Q^2 (r_+^2-Q^2) \alpha_2 \ell^2}{4 \pi r_+^7}-\frac{Q^4 \alpha_2^2 \ell^4}{8 \pi r_+^{11}}(r_+^2-Q^2)\\&+\frac{3 \beta_1 \ell^4}{28 \pi r_+^{11}}(109 Q^6-147 Q^4 r_+^2+73 Q^2 r_+^2-7 r_+^6)+\frac{\beta_2 Q^2 \ell^4}{4 \pi r_+^{11}}(7 Q^4-6 Q^2 r_+^2+2 r_+^4)\\&+\frac{\beta_3 Q^2 \ell^4}{4 \pi r_+^{11}}(7 Q^4-11 Q^2 r_+^2+4 r_+^4)+\mathcal{O}(\ell^6)\,.
\end{split}
\label{eq:temperature}
\end{equation}
Our next goal is the computation of the black hole entropy. According to the Iyer-Wald prescription \cite{Wald:1993nt,Iyer:1994ys}, the entropy $S$ of a black hole arising as a solution of a theory with Lagrangian density $\mathcal{L}$ is given by
\begin{equation}
S=-2 \pi \int_\Sigma \mathrm{vol}_\Sigma \frac{\delta \mathcal{L}}{\delta R_{\mu \nu \rho \sigma}} \epsilon_{\mu \nu} \epsilon_{\rho \sigma}\,,
\label{eq:iyerwald}
\end{equation}
where $\Sigma$ is the bifurcation surface of the horizon and $\mathrm{vol}_\Sigma$ its (induced) volume form. Similarly, $\epsilon_{\mu \nu}$ denotes the binormal to the $\Sigma$, which is nothing else than the components of the volume form in the orthogonal space to the horizon, and 
\begin{equation}
\dfrac{\delta \mathcal{L}}{\delta R_{\mu \nu \rho \sigma}}=\dfrac{\partial \mathcal{L}}{\partial R_{\mu \nu \rho \sigma}}-\nabla_{\alpha}\dfrac{\partial  \mathcal{L}}{\partial \nabla_{\alpha} R_{\mu \nu \rho \sigma}}+\dots
\end{equation}
is the functional derivative of the Lagrangian with respect to the Riemann tensor. From the spherical symmetry of the black hole horizon, it follows that we just have to work out the component $\dfrac{\delta \mathcal{L}}{\delta R_{t r t r }}$, which reads:
\begin{equation}
\begin{split}
\frac{\delta \mathcal{L}}{\delta R_{t r t r }}\Bigg|_{\Sigma}&=-\frac{1}{32 \pi}-\frac{\alpha_1 \ell^2 }{8\pi r_+^2} + \frac{\alpha_2 \ell^2 Q^2}{8 \pi r_+^4}+\frac{\alpha_1 \alpha_2 \ell^4 Q^2}{4\pi r_+^6}+\frac{\alpha_2^2 \ell^4 Q^2}{4\pi r_+^8}(-3 Q^2+2 r_+^2)\\&+\frac{\beta_1 \ell^4}{\pi r_+^8} \left (- \frac{1252 Q^4}{112}+\frac{519 Q^2 r_+^2}{56}-\frac{33 r_+^4}{16}\right)+\frac{\beta_2 \ell^4 Q^2}{8 \pi r_+^8}\left (-\frac{51}{4}Q^2+5 r_+^2 \right )\\&+\frac{\beta_3 \ell^4 Q^2}{16 \pi r_+^8}(-31Q^2+32 r_+^2)\,.
\end{split}
\end{equation}
After these computations, direct application of the Iyer-Wald formula \eqref{eq:iyerwald} yields 
\begin{align}\notag
S=&\pi r_+^2+4\pi \alpha_1 \ell^2 -\frac{2\pi Q^2 \alpha_2 \ell^2}{r_+^2}+\frac{12 \pi \beta_1 \ell^4 }{r_+^6}(r_+^2-2 Q^2)^2 +\frac{\pi Q^2 \beta_2 \ell^4}{r_+^6}(7 Q^2-4 r_+^2)\\
&+\frac{8 \pi Q^2 \beta_3 \ell^4}{r_+^6}(Q^2-r_+^2)+\mathcal{O}(\ell^6)\,.
\label{eq:entropy}
\end{align}
While the Gauss-Bonnet term, being topological, does not affect at all to the equations of motion of the theory, we observe that it does contribute to the corresponding black hole entropy by introducing a constant term which, in turn, does not have any influence on the first law of black hole thermodynamics.

Let us finally compute the electrostatic and magnetic potentials at the horizon, $\Phi(r_+)$ and  $\Psi(r_+)$. These are defined by
\begin{equation}
F=d (\Phi(r) dt)-\star d (\Psi(r)dt)\,,
\end{equation}
and comparing to \eqref{eq:fsbhsol}, we have that
\begin{equation}
\Phi(r)=q \chi(r) \, , \quad \Psi(r)=p \chi(r)\,, \quad \text{where}\quad  \chi'(r)=\frac{N(r)}{r^2}\,. 
\end{equation}
Imposing both $\Phi(r)$ and $\Psi(r)$ to vanish at infinity, and using the perturbative expression for $N(r)$ found in \eqref{eq:npertur}, we encounter
\begin{equation}
\begin{split}
\chi(r)&=\frac{1}{r}+\frac{Q^2 \alpha_2 \ell^2}{5r^5}+\frac{Q^2 \alpha_2^2 \ell^4}{14 r^9}(-63 Q^2+16(7M-3 r)r)+\frac{2 Q^2 \beta_2 \ell^4}{3 r^9}(-5 Q^2+3 M r)\\&-\frac{2 \beta_1 \ell^4}{7 r^9}(119 Q^4-156 M Q^2 r+54 M^2 r^2)-\frac{Q^2 \beta_3 \ell^4}{21 r^9}(175 Q^2+36 r(-7M+2r)))+\mathcal{O}(\ell^6)\,.
\end{split}
\end{equation}
Evaluation at the black horizon yields:
\begin{equation}
\begin{split}
\chi(r_+)&=\frac{1}{r_+}+\frac{Q^2 \alpha_2 \ell^2}{5r_+^5}-\frac{Q^2 \alpha_2^2 \ell^4}{14r_+^9}(7 Q^2-8 r_+^2)-\frac{\beta_1 \ell^4}{7 r_+^9}(109 Q^4-102 Q^2 r_+^2+27 r_+^4)\\&+\frac{Q^2 \beta_2 \ell^4}{3 r_+^9}(3 r_+^2-7 Q^2)+\frac{Q^2 \beta_3 \ell^4}{21 r_+^9}(-49 Q^2 +54 r_+^2)+\mathcal{O}(\ell^6)\,.
\label{eq:potentialchi}
\end{split}
\end{equation}
Making use of \req{eq:massrh}, \req{eq:entropy}, \req{eq:temperature} and \req{eq:potentialchi} we observe that the following identities hold:
\begin{equation}
\frac{\partial M}{\partial r_+}=T \frac{\partial S}{\partial r_+}\, , \quad \frac{1}{Q}\frac{\partial M}{\partial Q}=\frac{T}{Q}\frac{\partial S}{\partial Q}+ \chi(r_+)\, .
\label{eq:usefulident}
\end{equation}
From this, it immediately follows that the first law of black hole thermodynamics,
\begin{equation}
d M=T dS+ \Phi(r_+) dq+\Psi(r_+) dp\, ,
\end{equation}
is identically satisfied. 

\subsection{Constraints from the weak gravity conjecture}
The string swampland program \cite{Vafa:2005ui} aims at finding universal features of low-energy effective theories with a consistent UV completion, so that one can discard those that do not satisfy those properties. In this respect, one of the most successful proposals is that of the weak gravity conjecture (WGC) \cite{ArkaniHamed:2006dz}, which has recently motivated the study of higher-derivative corrections to charged extremal black holes \cite{Cheung:2018cwt,Hamada:2018dde,Bellazzini:2019xts,Charles:2019qqt,Loges:2019jzs,Cano:2019oma,Cano:2019ycn,Andriolo:2020lul,Loges:2020trf,Cano:2020qhy}.
\subsubsection*{Corrections to the charge-to-mass ratio}
Let us briefly review how the weak gravity conjecture can be used to constrain effective gravitational theories. A heuristic form of this conjecture states that all black holes, including extremal ones, should be able to decay. In order for (near-) extremal black holes to decay, it follows that there must exist a particle with a charge-to-mass ratio larger than the one of a extremal black hole --- otherwise the black hole cannot be discharged due to the extremality bound in GR: $M\ge |Q|$. However, in the presence of higher-derivative corrections, the extremal charge-to-mass ratio is not a constant but a function of the charge, and hence this statement depends on the size of the black hole. To see this, let us consider an extremal black hole with electric charge $Q$ and mass $M_{\rm ext}(Q)$, and let us assume that it discharges by emitting a particle with charge $q$ and mass $m$. The resulting black hole will have mass and charge $M'= M_{\rm ext}(Q)-m$, $Q'=Q-q$, but, on account of the extremality bound,  the mass must satisfy $M'\ge M_{\rm ext}(Q')$. Then, assuming that $q<<Q$, one can see that this condition is satisfied only if 
\begin{equation}\label{eq:mqratio}
\frac{m}{q}\le \frac{dM_{\rm ext}}{dQ}\, .
\end{equation}
This provides the bound on the particle's charge-to-mass ratio in order for the black hole to discharge. For large black holes $\frac{dM_{\rm ext}}{dQ}=1$ and hence it is enough to have a particle with $q/m\ge 1$. However, if $\frac{dM_{\rm ext}}{dQ}$ decreases as the black hole discharges, the bound \req{eq:mqratio} may be violated at certain point, and hence the evaporation is obstructed. A simple way to avoid this problem consists in demanding $\frac{dM_{\rm ext}}{dQ}$ to grow as $Q$ decreases, or in other words,

\begin{equation}\label{eq:wgc1}
\frac{d^2M_{\rm ext}}{dQ^2}\le 0\, .
\end{equation}

A more robust argument can be obtained from a slightly different form of the WGC which states that the decay of a black hole into a set of smaller black holes should be possible in terms of energy and charge conservation \cite{Hamada:2018dde}. This is like saying that smaller black holes play the role of the particle hypothesized by the WGC. Let us suppose that the initial black hole is extremal (or arbitrarily close to extremality), so that it has mass $M_{\rm ext}(Q)$, while the final black holes are not necessarily extremal and have charges $Q_1$, $Q_2$ with $Q_1+Q_2=Q$, and masses $M_1$, $M_2$. A necessary condition in order for this process to be admissible is that $M_{\rm ext}(Q_1+Q_2)\ge M_1+M_2$, which, upon use of the extremality bound $M_i\ge M_{\rm ext}(Q_i)$, yields the following constraint
\begin{equation}\label{eq:wgc2}
M_{\rm ext}(Q_1+Q_2)\ge M_{\rm ext}(Q_1)+M_{\rm ext}(Q_2)\, .
\end{equation}
Thus, the higher-derivative corrections to $M_{\rm ext}$ must be such that this condition is satisfied. 

When applied to our result \req{eq:Mext}, either condition \req{eq:wgc1} or \req{eq:wgc2} yields equivalent constraints. For instance, from \req{eq:wgc1} we get

\begin{equation}
-\frac{2\alpha_2\ell^2}{5 Q^3}-\frac{2\ell^4(3 \alpha_2^2+48\beta_1+7\beta_2+10 \beta_3)}{21 Q^5}+\mathcal{O}(\ell^6)\le 0\, .
\end{equation}
Obviously, in the regime where $Q>>\ell$ the first term is dominant and this implies that
\begin{equation}
\alpha_2\ge 0\, .
\label{eq:alpha2pos}
\end{equation}
Now, the second term only becomes relevant when $Q\sim \ell$, but this point marks the limit of applicability of the perturbative expansion, so it is not clear what constraint one should impose on the additional couplings. Still, one may argue that, if the coefficient of $1/Q^5$ is positive, then the bound may be eventually violated. In order to avoid this problem, it seems reasonable to impose as well the condition

\begin{equation}
3 \alpha_2^2+48\beta_1+7\beta_2+10 \beta_3\ge 0\, .
\label{eq:apos}
\end{equation}
In this way, we guarantee that the mild form of the WGC is satisfied. 

\subsubsection*{Positivity of the corrections to the entropy}
In another vein, it has been argued \cite{Cheung:2018cwt} that negative corrections to the mass of extremal black holes are in correlation with positive corrections to the black hole entropy. Although such connection has been proven in \cite{Goon:2019faz}, the relation is not complete. Strictly speaking, only the corrections to the near-extremal entropy are related to the corrections to the extremal mass, while the corrections to the extremal entropy are independent, as explained in \cite{Cano:2019ycn}.  Likewise, the corrections for neutral BHs are also independent.  On the other hand, the relation proven in \cite{Goon:2019faz} only applies for the leading order corrections, but it will probably not hold for the subleading ones. 
This motivates us to study the range of values of the couplings of the theory defined in \eqref{eq:EFT6} for which the corrections to the entropy are non-negative. We will demand that, for any charge and mass, the corrections of each order are non-negative independently, which is the strongest condition one may impose. 

We shall study the corrections in the non-extremal and extremal regimes. We begin by considering this latter case, since it is simpler. In fact, by replacing \req{eq:rext} in \eqref{eq:entropy} we find the following remarkably compact expression for the extremal entropy $S_{\mathrm{ext}}$:
\begin{equation}
S_{\mathrm{ext}}=\pi Q^2+2\pi(2\alpha_1-\alpha_2) \ell^2+\mathcal{O}(\ell^6)\,,
\end{equation}
which contains no $\ell^4$ corrections. These extremal corrections are non-negative whenever 
\begin{equation}
\alpha_1 \geq \frac{\alpha_2}{2}\,.
\label{eq:condext}
\end{equation}
Although the Gauss-Bonnet term is topological, we observe that demanding the corrections to extremal entropy to be non-negative imposes a bound on $\alpha_1$ --- in particular, it cannot vanish if $\alpha_2>0$. This condition did not appear when studying corrections to the extremal charge-to-mass ratio, since, as we anticipated, the corrections to the extremal black hole entropy are unrelated to those of the extremal charge-to-mass ratio \cite{Cano:2019ycn}.

Let us now study the entropy of non-extremal black holes. For that, we need to express first the black hole entropy in terms of the mass and charge, $S=S(M,Q)$. Note that back in \eqref{eq:entropy} we wrote the black hole entropy as a function of $r_+$ and $Q$, but we must bear in mind that the truly thermodynamic variables in which to express the entropy are the mass and charge. This is an important issue since the corrections at fixed $r_+$ are not the same as those at fixed $M$, the reason for this being that for a fixed mass $M$ the horizon radius $r_+$ is altered after the inclusion of corrections, and vice-versa. 
In terms of the extremality parameter $\zeta=\sqrt{1-\frac{Q^2}{M^2}}$, we find that the entropy $S_{\mathrm{ne}}$ in the non-extremal regime reads
\begin{equation}
\begin{split}
S_{\mathrm{ne}}&=\pi M^2 (1+\zeta)^2+4 \pi \ell^2 \alpha_1+\frac{2\pi \ell^2 \alpha_2(1-\zeta)^2}{5 \zeta (1+\zeta)}+\frac{a \pi \ell^4 (\zeta-1)^2(8\zeta+1)}{63 M^2 \zeta (1+\zeta)^4} \\&+\frac{ 8 \beta_1 \pi \ell^4 (3+4 \zeta) \zeta }{7 M^2 (1+\zeta)^4}+\frac{\pi (1-\zeta)^2 \ell^4 \alpha_2^2}{525 M^2 \zeta^3 (1+\zeta)^4} (-21-126 \zeta -185 \zeta^2+32 \zeta^3)+\mathcal{O}(\ell^6)\,,
\end{split}
\end{equation}
where we have defined $a=48 \beta_1+7 \beta_2+10 \beta_3$. We recall that this expression is valid for $\zeta\gtrsim \ell/M$, which in practice can be very small. 
Regarding the $\mathcal{O}(\ell^2)$ corrections, we see that the term with $\alpha_2$ is dominant for small $\zeta$, and therefore positivity demands that
\begin{equation}
\alpha_2\ge 0\, ,
\end{equation}
which is the same condition as the one from the corrections to the extremality bound. On the other hand, \req{eq:condext} ensures that the $\mathcal{O}(\ell^2)$ corrections are positive for any other value of $\zeta$. 
If we now demand that the  $\mathcal{O}(\ell^4)$ corrections be non-negative independently, we see that the following constraints are obtained,
\begin{equation}\label{eq:consS2}
\alpha_2=0\, ,\quad 48 \beta_1+7 \beta_2+10 \beta_3\ge 0\, ,\quad \beta_1\ge 0\, .
\end{equation}
The first two are obtained by examining the corrections for small $\zeta$ while the bound on $\beta_1$ is obtained in the opposite limit, $\zeta=1$. Note that these constraints are quite strong since they imply the vanishing of $\alpha_2$, so one may wonder if imposing the non-negativity of entropy corrections at each order is a well-motivated bound.  In any case, note that we recover \eqref{eq:apos} with $\alpha_2=0$.

\section{Conclusions}\label{sec:conclusions}

In this work, we have studied higher-order extensions of Einstein-Maxwell theory which are invariant under electromagnetic duality rotations. We started by considering a general higher-derivative theory of gravity and electromagnetism and we obtained the necessary and sufficient conditions for theories up to eight derivatives to preserve electromagnetic duality at a perturbative level. It would be interesting to extend this result to all orders in the derivative expansion, but we leave this task for the future. Then we used these results to derive the most general parity- and duality-invariant theory up to eight derivatives.

We also studied in some detail a particular class of theories: those with Lagrangian $\mathcal{L}=R-\chi^{\mu \nu \rho \sigma} F_{\mu \nu} F_{\rho \sigma}$, where the susceptibility tensor $\chi^{\mu \nu \rho \sigma}$ only depends on the curvature tensor and the metric. We provided the most general form for $\chi^{\mu \nu \rho \sigma}$ yielding a duality-invariant theory and we wrote the corresponding set of equations of motions with manifest $\mathrm{SO}(2)$ invariance. We observed that, generically, these theories consist of an infinite tower of higher-derivative terms, but it would be interesting to study if there exist particular theories or situations under which this infinite series can be explicitly summed to obtain a complete non-perturbative duality-invariant theory. Research in this direction is underway \cite{CMdualityBH}.

Next we studied the effect of field redefinitions on duality-invariant theories, which led us to a remarkable simplification of those actions. We showed that, up to six derivatives, one can always remove all the higher-order terms with explicit Maxwell field strengths, leaving one with a higher-derivative gravity minimally coupled to the Maxwell Lagrangian. In other words, this result implies that, in a duality-invariant theory, higher-derivative corrections can always be chosen in a scheme such that they do not modify the Maxwell Lagrangian at all.  We argued that this phenomenon should take place for any theory with any number of derivatives, but so far this claim remains as a conjecture. Appealing open questions are those of providing a proof for the conjecture or showing that any theory is equivalent via metric redefinitions to Maxwell theory coupled to a higher-order gravity at a non-perturbative level as well. 

In this context, we wrote the most general six-derivative duality-invariant action after the use of metric redefinitions, noticing that the number of higher-order operators can be effectively reduced to four, plus a topological one. We studied the charged, static and spherically symmetric black hole solutions of this theory and computed their thermodynamic properties --- entropy, temperature and electromagnetic potentials at the horizon --- allowing us to check explicitly that the first law of thermodynamics holds. 

Using these results, we obtained additional constraints on the higher-derivative couplings by applying the recently proposed mild form of the weak gravity conjecture \cite{Hamada:2018dde,Bellazzini:2019xts}.
According to this conjecture, the corrections to the charge-to-mass ratio of extremal black holes in any consistent theory of quantum gravity should be non-negative, thus allowing the decay of extremal black holes and evading the existence of remnants. This has also been related to the non-negativity of entropy corrections \cite{Cheung:2018cwt,Goon:2019faz}, although, as we discussed, the connection is not complete. We determined the constraints coming from these conditions not only for the leading higher-derivative corrections but also for the subleading ones. Demanding that the subleading corrections to the entropy remain non-negative leads to especially strong constraints --- see \req{eq:consS2} ---, and therefore it would be interesting to understand whether imposing such condition is justified, or if, on the contrary, the mild form of the WGC should only be applied to the leading corrections.

Let us close this work by commenting on future directions. As we have already remarked, it would be interesting to extend the analysis of this paper to arbitrary orders in the derivative expansion and, if possible, to obtain examples of exactly invariant theories at the non-perturbative level.
On the other hand, in our analysis we left out the terms that depend on derivatives of the field strength, because, as we argued, such terms cannot appear in a truncation of a theory that is exactly invariant under duality. However, they can preserve duality in a perturbative sense  \cite{Chemissany:2011yv}, so it may have certain interest to classify such theories as well. It would also be interesting to generalize the results in this work to the case of $n$ vector fields coupled to scalar fields. In those cases, the duality group is a subgroup of $\mathrm{Sp}(2n,\mathbb{R})$, and although some of these Lagrangians have been studied in detail \cite{Gaillard:1981rj,Gaillard:1997zr,Gaillard:1997rt} and are relevant in supergravity and string theory, it appears that couplings between the field strength and the curvature have not been considered so far. 
Finally, it is also possible to study duality-invariant theories in higher dimensions, for instance by considering $p-$forms in $D=2p+2$ dimensions.

\section*{Acknowledgments}
We would like to thank Tom\'as Ort\'in and Carlos S. Shahbazi for useful discussions and comments. The work of PAC is supported by a postdoctoral fellowship from the Research Foundation - Flanders (FWO grant 12ZH121N). The work of \'AM is funded by the Spanish FPU Grant No.  FPU17/04964 and by the Deutscher Akademischer Austauschdienst (DAAD), through the Short-Term Research Grant No. 91791300. \'AM was further supported by the MCIU/AEI/FEDER UE grant PGC2018-095205-B-I00 and by the ``Centro de Excelencia Severo Ochoa'' Program grant SEV-2016-0597.

\appendix

\section{Invariance of Einstein's equations in theories with derivatives of the curvature}\label{app:inv}

In this appendix we are going to show that the invariance of the constitutive relation \eqref{RotCons} under duality transformations implies the invariance of the Einstein's equations also for those theories with explicit covariant derivatives of the curvature. For that, we shall follow an analogous reasoning to that of Subsection \ref{subsec:inveineq}, where we showed the invariance of the Einstein's equations under duality rotations once that of \eqref{RotCons} is guaranteed. 

If we consider a generic theory with arbitrary dependence on the covariant derivatives of the Riemann curvature tensor, so that $\mathcal{L}=\mathcal{L}(g^{\mu \nu}, F_{\mu \nu} , R_{\mu \nu \rho \sigma},  \nabla_\alpha R_{\mu \nu \rho \sigma},\dots)$, the main difficulty we encounter is that the tensor $\mathcal{E}_{\mu \nu}$, defined back in Eq. \eqref{eq:Emunu1},  takes a much more complicated form to that given at \eqref{EmunuEinst}. However, we can make use of many computations and results presented in Subsection \ref{subsec:inveineq} for algebraic theories to achieve our goal. In fact, the arguments used for algebraic theories are valid in this general case up to Eq.~\req{eq:emunuexpanmejor}. Thus, our task will be to show the validity of the equations \eqref{eq:firstidentityeins} and \eqref{eq:secondidentityeins} also for these general theories so that the invariance of the Einstein's equations will be guaranteed. We rewrite here those equations for the benefit of the reader: 
\begin{eqnarray}
\label{eq:appfirstidentityeins}
& &\mathcal{E}^{\rm IH (6)}_{\mu\nu}(T)+	\frac{1}{4}\frac{\delta \mathcal{E}^{(4)}_{\mu\nu} }{\delta F_{\alpha\beta}}(T)\circ \frac{\partial \mathcal{L}_{(4)}}{\partial \tensor{F}{^{\alpha\beta}}}=0\, ,\\ \nonumber
 & & \mathcal{E}^{\rm IH(8)}_{\mu\nu}(T) +\frac{1}{4}\frac{\delta \mathcal{E}^{(4)}_{\mu\nu}}{\delta F_{\alpha\beta}} (T) \circ \frac{\partial \mathcal{L}_{(6)}}{\partial F_{\alpha \beta}}-\frac{1}{16}\frac{\delta^2 \mathcal{E}^{(4)}_{\mu\nu}}{\delta F_{\alpha\beta} \delta F_{\rho \sigma}}(T) \circ  \frac{\partial \mathcal{L}_{(4)}}{\partial F^{\alpha \beta} } \circ \frac{\partial \mathcal{L}_{(4)}}{\partial F^{\rho \sigma} }\\& &\label{eq:appsecondidentityeins}+\frac{1}{4}\frac{\delta \mathcal{E}^{\rm H(6)}_{\mu\nu}}{\delta F_{\alpha\beta}} (T) \circ \frac{\partial \mathcal{L}_{(4)}}{\partial F^{\alpha \beta}}=0\,.
\end{eqnarray}

Let us first of all find the form of the equations of motion for theories with dependence on derivatives of the curvature.
For that, let us define
\begin{equation}
P{}_{\nabla^n}{}^{\alpha_1 \dots \alpha_n \mu \nu \rho \sigma}=\frac{\partial \mathcal{L}}{\partial \nabla_{\alpha_1 \dots \alpha_n} R_{\mu \nu \rho \sigma}}\,,
\end{equation}
 which enjoys the same symmetries as $\nabla_{\alpha_1 \dots \alpha_n} R_{\mu \nu \rho \sigma}$, which is a short-hand notation for $\nabla_{\alpha_1} \dots \nabla_{\alpha_n} R_{\mu \nu \rho \sigma}$. Consequently, the metric variation of the corresponding action $S[g,A]$ associated to $\mathcal{L}(g^{\mu \nu}, F_{\mu \nu} , R_{\mu \nu \rho \sigma},  \nabla_\alpha R_{\mu \nu \rho \sigma},\dots)$ takes the form:
\begin{equation}
\begin{split}
\delta S [g,A](\delta g^{\mu \nu},0)=\frac{1}{16 \pi G}\int_M d^4 x \sqrt{\vert g \vert}& \left\lbrace -\frac{1}{2}g_{\mu \nu} \delta g^{\mu \nu} \mathcal{L}+\frac{\partial \mathcal{L}}{\partial g^{\mu \nu}} \delta g^{\mu \nu}\right. \\& \left. +\sum_{n=0}^\infty P{}_{\nabla^n}{}^{\alpha_1 \dots \alpha_n \mu \nu \rho\sigma } \delta \nabla_{\alpha_1 \dots \alpha_n}  R_{\mu \nu \rho \sigma} \right\rbrace\,.
\end{split}
\label{eq:varactiongeneral}
\end{equation}
If $\xi \in \mathfrak{X}(M)$ denotes an arbitrary vector field, we can write that the Lie derivative $L_\xi \mathcal{L}$ in two different ways:
\begin{eqnarray}
L_\xi \mathcal{L}&=&\xi^\kappa \nabla_\kappa \mathcal{L}=\xi^\kappa \sum_{n=0}^\infty P{}_{\nabla^n}{}^{\alpha_1 \dots \alpha_n \mu \nu \rho \sigma} \nabla_\kappa \nabla_{\alpha_1}\dots \nabla_{\alpha_n} R_{\mu \nu \rho \sigma} +\xi^\kappa \frac{\partial \mathcal{L}}{\partial F^{\alpha \beta}} \nabla_\kappa F^{\alpha \beta}\,, \\
L_\xi \mathcal{L}&=& \sum_{n=0}^\infty P{}_{\nabla^n}{}^{\alpha_1 \dots \alpha_n \mu \nu \rho \sigma} L_\xi \nabla_{\alpha_1\dots \alpha_n} R_{\mu \nu \rho \sigma} +\frac{\partial \mathcal{L}}{\partial F_{\alpha \beta}} L_\xi F_{\alpha \beta}+\frac{\partial \mathcal{L}}{\partial g_{\alpha \beta}} L_\xi g_{\alpha \beta}\,.
\end{eqnarray}
On the other hand, we have the following identities:
\begin{eqnarray}
\nonumber
P{}_{\nabla^n}{}^{\alpha_1 \dots \alpha_n \mu \nu \rho \sigma} L_\xi \nabla_{\alpha_1 \dots \alpha_n} R_{\mu \nu \rho \sigma}&=&\xi^\kappa  P{}_{\nabla^n}{}^{\alpha_1 \dots \alpha_n \mu \nu \rho \sigma} \nabla_{\kappa \alpha_1\dots \alpha_n} R_{\mu \nu \rho \sigma} \\ & +&\,4 (\nabla_\kappa \xi_\gamma) P{}_{\nabla^n}{}^{\alpha_1 \dots \alpha_n \kappa \nu \rho \sigma} \nabla_{\alpha_1\dots  \alpha_n} R^\gamma{}_{\nu \rho \sigma}\\ \nonumber & + &\sum_{i=0}^n  (\nabla_{\alpha_i} \xi_{\beta_i}) P{}_{\nabla^n}{}^{\alpha_1 \dots \alpha_i \dots \alpha_n \mu \nu \rho \sigma} \nabla_{\alpha_1 \dots }{}^{\beta_i}{}_{\dots \alpha_n} R_{\mu \nu \rho \sigma}\, , \\
L_\xi g_{\alpha \beta}&=&2 \nabla_{(\alpha} \xi_{\beta)}\,,\\
\frac{\partial L}{\partial F_{\alpha \beta} }L_\xi F_{\alpha \beta}&=&\xi^\mu \nabla_\mu F_{\alpha \beta} \frac{\partial L}{\partial F_{\alpha \beta}}+2 \nabla_\alpha \xi^\mu F_{\mu \beta} \frac{\partial \mathcal{L}}{\partial F_{\alpha \beta}}\,.
\end{eqnarray} 
Consequently, we learn that
\begin{equation}
\begin{split}
\frac{\partial \mathcal{L}}{\partial g_{\mu \nu}}&=-2 P{}_{\nabla^n}{}^{\alpha_1 \dots \alpha_n \mu \lambda \rho \sigma} \nabla_{\alpha_1\dots  \alpha_n} R^\nu{}_{\lambda \rho \sigma}-\frac{\partial \mathcal{L}}{\partial F_{\mu \rho}}F^{\nu \rho}\\&-\frac{1}{2}\sum_{i=1}^n P{}_{\nabla^n}{}^{\alpha_1 \dots \hat{\mu} \dots \alpha_n \lambda \kappa \rho \sigma} \nabla_{\alpha_1 \dots }{}^{\hat{\nu}}{}_{\dots \alpha_n} R_{\lambda \kappa \rho \sigma}\, ,
\label{eq:dlpartialg}
\end{split}
\end{equation}
where the hats over the free indices $\mu$ and $\nu$ denote that they replace the indices $\alpha_i$ in the $i$-th position. Taking into account that, up to total derivatives,
\begin{equation}
\begin{split}
P{}_{\nabla^n}{}^{\alpha_1 \dots \alpha_n \mu \nu \rho\sigma } \delta \nabla_{\alpha_1 \dots \alpha_n}  R_{\mu \nu \rho \sigma}&=(-1)^{n+1} \nabla_{\alpha_n \dots \alpha_1} P{}_{\nabla^n}{}^{\alpha_1 \dots \alpha_n}{}_\mu{}^{\nu \rho\sigma }R_{\beta \nu \rho \sigma} \delta g^{\mu \beta}\\&+2 (-1)^n\nabla^\sigma \nabla^\beta \nabla_{\alpha_n \dots \alpha_1} P{}_{\nabla^n}{}^{\alpha_1 \dots \alpha_n}{}_{\mu\sigma \nu \beta } \delta g^{\mu \nu}\,,
\end{split}
\end{equation}
we find that
\begin{equation}
\begin{split}
\mathcal{E}_{\mu \nu}&=\sum_{n=0}^{n_{\mathrm{max}}} \bigg [ 2 (-1)^{n+1}  \nabla^\sigma \nabla^\beta \nabla_{\alpha_n \dots \alpha_1} P_{\nabla^n}{}^{\alpha_1 \dots \alpha_n}{}_{(\mu | \sigma | \nu) \beta}+(-1)^n \nabla_{\alpha_n \dots \alpha_1 } P_{\nabla^n}{}^{\alpha_1 \dots \alpha_n}{}_{(\mu }{}^{\rho \sigma \gamma} R_{\nu) \rho \sigma \gamma}\\&- 2 P_{\nabla^n}{}^{\alpha_1 \dots \alpha_n}{}_{(\mu|}{}^{\lambda \rho \sigma} \nabla_{\alpha_1 \dots \alpha_n} R_{|\nu) \lambda \rho \sigma}-\frac{1}{2}\sum_{i=1}^n P{}_{\nabla^n}{}^{\alpha_1 \dots}{}_{(\hat{\mu}|}{}^{\dots \alpha_n \lambda \kappa \rho \sigma} \nabla_{\alpha_1 \dots |\hat{\nu}) \dots \alpha_n} R_{\lambda \kappa \rho \sigma} \bigg]\\& + \frac{1}{2} g_{\mu \nu} \left ( \mathcal{L}-\frac{1}{2} F_{\alpha \beta} \frac{\partial \mathcal{L}}{\partial F_{\alpha \beta}} \right ) \,,
\end{split}
\label{eq:emunucovariant}
\end{equation}
where $n_{\mathrm{max}}$ is the maximum number of explicit covariant derivatives appearing in the action. 

Let us remark at this point that, as in the case of algebraic theories, the equations $\mathcal{E}^{{\rm H} (2n) }_{\mu\nu}$ associated to the homogeneous Lagrangians $\mathcal{L}^{\rm H}_{(2n)}$ only depend on $F_{\mu\nu}$ through the Maxwell stress tensor $T_{\mu\nu}$. This follows from the fact that for every monomial we have $F_{\alpha \beta} \frac{\partial \mathcal{L}_i}{\partial F_{\alpha \beta}}\propto\mathcal{L}_i$. On the other hand, if $ \mathcal{L}$ is a function of $T_{\mu\nu}$ so are the various $ P_{\nabla^n}$ tensors. Thus we can in fact write $\mathcal{E}^{{\rm H} (2n)}_{\mu\nu}(T)$, so that we can apply Eq.~\req{eq:emunuexpanmejor}.


Finally, in order to show \req{eq:appfirstidentityeins} and \req{eq:appsecondidentityeins}, let us note the following formula, which generalizes \eqref{eq:propnabladerfunc} for an arbitrary number of covariant derivatives:
\begin{equation}
\frac{\delta \nabla_{\mu_1 \dots \mu_n} \mathcal{Q}_{\nu_1 \dots \nu_n }}{\delta F_{\alpha \beta}} \circ \mathcal{A}_{\alpha \beta} =\nabla_{\mu_1 \dots \mu_n} \left ( \frac{\delta  \mathcal{Q}_{\nu_1 \dots \nu_n} }{\delta F_{\alpha \beta}} \circ \mathcal{A}_{\alpha \beta} \right )\,,
\label{eq:inveqdemos}
\end{equation}
where $\mathcal{Q}_{\nu_1 \dots \nu_n }$ is any tensor with dependence on $F_{\alpha \beta}$ and its covariant derivatives and where $\mathcal{A}_{\alpha \beta}$ is an arbitrary antisymmetric tensor. Using this formula, and noting the structure of \eqref{eq:emunucovariant}, we check after some computations that both \eqref{eq:appfirstidentityeins} and \eqref{eq:appsecondidentityeins} hold. For the sake of clarity, let us illustrate this fact more explicitly for Eq.~\eqref{eq:appfirstidentityeins}. By use of \eqref{eq:inveqdemos}, we have that
\begin{equation}
\begin{split}
\frac{\delta \mathcal{E}^{(4)}_{\mu\nu} }{\delta F_{\alpha\beta}}(T)\circ \frac{\partial \mathcal{L}_{(4)}}{\partial \tensor{F}{^{\alpha\beta}}}&=\sum_{n=0}^{n_{\mathrm{max}}} \bigg [ 2 (-1)^{n+1}  \nabla^\sigma \nabla^\beta \nabla_{\alpha_n \dots \alpha_1} \hat{P}_{\nabla^n}{}^{\alpha_1 \dots \alpha_n}{}_{(\mu | \sigma | \nu) \beta}\\&+(-1)^n \nabla_{\alpha_n \dots \alpha_1 } \hat{P}_{\nabla^n}{}^{\alpha_1 \dots \alpha_n}{}_{(\mu }{}^{\rho \sigma \gamma} R_{\nu) \rho \sigma \gamma}- 2 \hat{P}_{\nabla^n}{}^{\alpha_1 \dots \alpha_n}{}_{(\mu|}{}^{\lambda \rho \sigma} \nabla_{\alpha_1 \dots \alpha_n} R_{|\nu) \lambda \rho \sigma}\\&-\frac{1}{2}\sum_{i=1}^n \hat{P}{}_{\nabla^n}{}^{\alpha_1 \dots}{}_{(\hat{\mu}|}{}^{\dots \alpha_n \lambda \kappa \rho \sigma} \nabla_{\alpha_1 \dots |\hat{\nu}) \dots \alpha_n} R_{\lambda \kappa \rho \sigma} \bigg] \\&+ \frac{1}{2} g_{\mu \nu} \frac{\partial \mathcal{L}_{(4)}}{\partial F^{\alpha \beta}}  \frac{\partial}{\partial F_{\alpha \beta}} \left ( \mathcal{L}_{(4)}-\frac{1}{2} F_{\rho \sigma} \frac{\partial \mathcal{L}_{(4)}}{\partial F_{\rho \sigma}} \right ) \,,
\end{split}
\label{eq:appemunucovariant}
\end{equation}
where we have defined 
\begin{equation}
\hat{P}_{\nabla^n}{}^{\alpha_1 \dots \alpha_n \mu \nu \rho \sigma}=\frac{\partial P_{\nabla^n}^{(4)\,\alpha_1 \dots \alpha_n \mu \nu \rho \sigma}}{ \partial F^{\lambda \kappa}} \frac{\partial \mathcal{L}_{(4)}}{\partial F_{\lambda \kappa}}\,.
\end{equation}
We identify in \eqref{eq:appemunucovariant} precisely the term $-4 \mathcal{E}^{\rm IH (6)}_{\mu\nu}(T)$, so we conclude. The proof of \eqref{eq:appsecondidentityeins} goes along similar lines.

\bibliographystyle{JHEP}
\bibliography{Biblio}

\end{document}